\documentclass{aa}
\usepackage{graphics,epsfig,astron,amsmath,amssymb,amstext}

\def\d {\mathrm{d}}
\def\Msun{\hbox{$M_{\odot}$}}
\def\tSW{\hbox{${t_{\rm SW}}$}}

\def\tend{\hbox{${t_{\rm end}}$}}

\begin{document}

\thesaurus{12(02.01.1; 02.14.1; 09.19.2; 10.01.1)}

\title{Spallative Nucleosynthesis in Supernova Remnants}
  
\subtitle{II. Time-dependent numerical results}

\author{Etienne Parizot \and Luke Drury}

\offprints{E.~Parizot}

\institute{Dublin Institute for Advanced Studies, 5 Merrion Square, 
Dublin 2, Ireland\\ e-mail: parizot@cp.dias.ie; ld@cp.dias.ie}

\date{(accepted for publication in A\&A)}

\maketitle

\begin{abstract}
We calculate the spallative production of light elements associated 
with the explosion of an isolated supernova in the interstellar 
medium, using a time-dependent model taking into account the dilution 
of the ejected enriched material and the adiabatic energy losses.  We 
first derive the injection function of energetic particles (EPs) 
accelerated at both the forward and the reverse shock, as a function 
of time.  Then we calculate the Be yields obtained in both cases and 
compare them to the value implied by the observational data for 
metal-poor stars in the halo of our Galaxy, using both O and Fe data.  
We find that none of the processes investigated here can account for 
the amount of Be found in these stars, which confirms the analytical 
results of Parizot and Drury (1999).  We finally analyze the 
consequences of these results for Galactic chemical evolution, and 
suggest that a model involving superbubbles might alleviate the 
energetics problem in a quite natural way.

\keywords{Acceleration of particles; Nuclear reactions, 
nucleosynthesis; ISM: supernova remnants; Galaxy: abundances}

\end{abstract}

\section{Introduction}
\label{Introduction}

The class of light elements, namely Li, Be and B, sets itself apart 
from any other by its interstellar origin (except for part of the 
$^{7}$Li, produced in the Big Bang ages, and perhaps part of the 
$^{11}$B, produced in supernova (SN) explosions by 
neutrino-spallation).  Concentrating on the most representative 
isotope, the abundance of $^{9}$Be in stars of increasing metallicity 
can be regarded as the witness and tracer of the nuclear spallation 
efficiency during Galactic chemical evolution.  Indeed, virtually 
every atom of Be observed in the atmosphere of stars must have been 
produced by the spallation of a larger nucleus, most probably C or O, 
induced by the interaction of energetic particles (EPs) with the 
interstellar medium (ISM).

Since the first measurement of Be in a very metal-poor star at the 
beginning of the decade (Gilmore et~al.  1991), increasing evidence 
has been gathered showing that the abundance of Be and B in the early 
Galaxy (until the ambient metallicity is 10\% that of the sun, say) 
kept increasing jointly and linearly with ordinary metallicity 
tracers, such as Fe or O, as if they were actually primary elements 
(Duncan et al.  1992,1997; Edvardsson et al.  1994; Gilmore et al.  
1992; Kiselman \& Carlsson 1996; Molaro et al.  1997; Ryan et al.  
1994).  Now they are not, since as we just recalled C and O nuclei 
have to be produced first in order that they can be spalled by EPs 
into light elements.  The observations therefore suggest that some 
process must act to ensure that, on average, an equal amount of Be is 
synthesized each time a given mass of Fe or O is ejected into the ISM. 
It should be clear, however, that this statement relies on the 
assumption that the abundances of O and Fe are proportional to one 
another, at least during the early evolution stages in which we are 
interested here.

This assumption has long been used with high confidence level based on 
both theoretical and observational arguments, but new observations 
seem to contradict it dramatically (Israelian et~al.  1998; Boesgaard 
et~al.  1998).  Although an independent confirmation of these 
observations would be welcome, they have recently been used to 
reappraise the alleged `primary behavior' of $^{6}$LiBeB Galactic 
evolution (Fields and Olive, 1999).  Indeed, if the O/Fe abundance 
ratio is not constant but actually decreases with metallicity, then 
the observed approximate constancy of the Be/Fe ratio implies an 
increasing Be/O ratio.  Fields and Olive (1999) find a Be--O 
logarithmic slope in the range 1.3--1.8, which seems to contradict 
both the primary scenario (slope 1) and the secondary scenario (slope 
2), in which the spallation reactions producing the light elements are 
induced by standard Galactic cosmic rays (GCRs) accelerated out of the 
ISM. However, the current lack of Be and O abundance measurements in 
the same very metal-poor stars (with [O/H] $= 10^{-3}$, say) makes the 
data marginally compatible, within error bars, with both scenarii.

While the situation should be soon clarified, notably by the 
accumulation of data at lower metallicity and independent measurements 
of Be, B, O and Fe in the same set of halo stars, we (Parizot and 
Drury, 1999; Paper~I) choose to investigate the Be production in the 
ISM from the other direction, i.e calculate the Be yield associated 
with the explosion of an isolated supernova (SN) in the ISM, according 
to current knowledge about supernova remnant (SNR) evolution and 
standard shock acceleration, and compare this Be yield with the value 
required to explain the observed Be/Fe ratio in metal-poor stars.  We 
identified two different mechanisms leading naturally to a primary 
evolution of Be in the early Galaxy.  In the first mechanism, 
particles from the ambient ISM (i.e.  metal-poor) are accelerated at 
the forward shock of the SN and confined within the SNR until the end 
of the Sedov-like evolution phase.  There, they interact with the 
freshly synthesized C and O nuclei, and therefore produce Be by 
spallation at a much higher rate than in the (secondary) GCR 
nucleosynthesis scenario in which they merely interact with the 
ambient, metal-poor ISM. In the second mechanism, particles from the 
enriched SN ejecta are accelerated at the reverse shock and again 
confined within the SNR during Sedov-like phase, where they suffer 
adiabatic losses through which they lose between 30\% and 70\% of 
their initial
energy, depending on the ambient density.  After the end of the 
Sedov-like phase, these particles diffuse out in the ISM where the 
energetic C and O nuclei can be spalled by the H and He atoms at rest 
in the Galaxy.

We have shown in Paper~I, through approximate analytical calculations, 
that the total Be yield obtained by processes~1 and~2 depends on the 
ambient density, and that this third mechanism is actually the most 
efficient (for light element production) in most cases, though not 
efficient enough to account for the observed Be/Fe ratio of $\sim 
1.6\,10^{-6}$.  If each SN ejects on average 0.1~\Msun~of Fe in the 
ISM, then the average Be yield per SN must be $\sim 4\,10^{48}$ atoms 
(cf.  Ramaty et al.  1997), which exceeds even our most optimistic 
calculated yields by about one order of magnitude.  We concluded that 
another mechanism or source of energy should be invoked, and argued 
that a model based on superbubble acceleration (involving the 
collective effect of SNe rather than individual SN shock acceleration) 
is a quite natural and promising candidate.  In this paper, we confirm 
the results of Paper~I by performing time-dependent numerical 
calculations, and discuss in more details their implications for 
Galactic chemical evolution scenarii.  The reader is referred to 
Paper~I for a more detailed description of the mechanisms considered 
here, and a discussion of their motivation and theoretical 
justification.

\section{Why we need to do time-dependent calculations}
\label{NonStationarity}

We intend to calculate the Li, Be and B (LiBeB) production induced by 
the interaction of energetic particles within a SNR. We shall first 
consider the fate of the particles accelerated out of the ambient, 
zero metallicity ISM entering the forward shock created by a SN 
explosion (process~1), and then turn to the acceleration of particles 
from the SN ejecta at the reverse shock, on a very short time scale 
around the so-called sweep-up time, $\tSW$ (process~2).  It turns out 
that both of these processes are highly non-stationary, for a number 
of reasons which we now review.

\subsection{EPs accelerated at the forward shock}

Considering first process~1, we expect that the particle injection 
power be more or less proportional to the power of the shock, which is 
a decreasing function of time as the SNR evolves.  Therefore the 
injection rate of the EPs is not constant, and no steady-state 
distribution function of the EPs within the SNR is ever reached.  If 
everything else was constant in the problem, we could however 
calculate the total energy injected in the form of EPs during the 
whole process, and multiply it by the steady-state spallation 
efficiency (defined as the `number of nuclei synthesized per erg 
injected'), evaluated from standard steady-state calculations.  This 
would provide us with the total spallation yields (i.e.  the time 
integral of the spallation rates), which are the only observationally 
relevant quantities.  This, however, cannot be done in the case we are 
considering, because the chemical composition of the target, namely 
the interior of the SNR, is also evolving during the expansion.

Indeed, as more and more metal-poor material is swept-up from the ISM 
by the shock, the metal-rich SN ejecta suffer stronger and stronger 
dilution, which makes the spallation of C and O less and less 
efficient.  As a consequence, even though we can evaluate the total 
energy eventually imparted to EPs, we cannot deduce the spallation 
yields from it because we don't know what composition to choose for 
the target.  Again, if this was the only non stationary feature in the 
process, we could still calculate the average target composition and 
compute the spallation yields from it.  But since both the EP 
injection rate \emph{and} the target composition are functions of 
time, steady-state models cannot be used in any consistent way, and a 
fully time-dependent calculation is required.

Qualitatively, it is easy to show that the yields which we obtain by 
integrating the time-dependent spallation rates must be appreciably 
higher than those derived from steady-state estimates using the total 
energy injected in the form of EPs and the (constant) mean target 
composition.  Indeed, the latter amounts to assuming that the 
injection rate is also constant and equal to the average power of the 
EPs (i.e.  the total energy divided by the duration of the process).  
However, in the time-dependent model, we take advantage of the fact 
that the EP power is higher at the beginning, when the target 
composition is richer in C and O. In other words, the spallation 
efficiency is higher when the EP fluxes are higher too, and 
conversely, less energy is imparted to the less efficient EPs 
accelerated towards the end of the process.

In addition to the sources of non-stationarity just mentioned 
(time-dependent injection and dilution of the ejecta), we also have to 
take into account the adiabatic losses suffered by the EPs as they 
wander inside the expanding volume of the SNR. Now these adiabatic 
losses are essentially function of time, becoming smaller and smaller 
as the SNR expands and the shock velocity gets lower.  This again can 
only be taken into account in a time-dependent model.

\subsection{EPs accelerated at the reverse shock}

Coming now to the case of process~2, where particles from the enriched 
material ejected by the supernova are accelerated at the reverse 
shock, it is clear that the dilution effect mentioned above does not 
have any significant influence anymore.  Indeed, the light element 
production is now dominated by the spallation of energetic C and O 
nuclei interacting with ambient H and He, instead of ambient C and O 
interacting with energetic H and He nuclei in the case of process~1 
(see Figs.~\ref{SNR2_f1} and~\ref{SNR2_f4}).  The abundance of (non 
energetic) C and O in the target has therefore only a negligible 
influence, since these nuclei hardly contribute to the spallation 
yields.  Nevertheless, the time dependence of the EP injection and the 
adiabatic losses still have to be taken into account, which is enough 
to make time-dependent calculations indispensable.

As has been argued in Paper~I, the curve representing the power in the 
reverse shock, $\mathcal{P}_{\mathrm{in}}$, as a function of time 
strongly peaks around the sweep-up time, $\tSW$, which is defined as 
usual as the time at which the swept-up mass is equal to the ejected 
mass.  This also approximately marks the end of the free expansion 
phase and the beginning of the adiabatic (or Sedov-like) phase.  In 
the absence of a motivated prescription for the reverse shock power 
function, $\mathcal{P}_{\mathrm{in}}(t)$, and on the understanding 
that its time scale is short as compared to the energy loss time 
scale, we shall consider below the injection of EPs as instantaneous 
in the case of process~2.  We thus just could not be further away from 
a steady-state.  However, as above, were this the only non-stationary 
feature of the process, we could still obtain the integrated 
spallation yields from steady-state calculations by merely multiplying 
the total energy injected in the form of EPs by the spallation 
efficiency, which in this case is almost independent of the target 
composition.  Unfortunately, as already indicated, the adiabatic 
losses are also a function of time, and will therefore cause the 
aforementioned spallation efficiency to vary as the process goes on.

On the other hand, once the EPs leave the SNR at the end of the 
Sedov-like phase to interact with the surrounding ISM, they will only 
suffer the usual Coulombian losses, which are essentially independent 
of time.  The above argument therefore does not apply anymore and this 
final part of process~2 (occuring outside the remnant) could be worked 
out with purely steady-state machinery.  This is in fact what we did 
in Paper~I (see its Sect.~4), in our study of what we then called 
process~3.  Here, however, we shall not distinguish between the part 
of the process occuring inside the SNR, and the part occuring outside 
(former process~3), because our time-dependent numerical model allows 
us to treat both on the same footing.  In particular, we obtain in 
this way not only the total LiBeB yields, but also their production 
rates as a function of time, whose time integral can be sucessfully 
checked to be equal to the steady-state yields.

\section{Description of the theoretical and numerical model}

It has been shown in the previous section that the spallative 
production of light elements associated with the explosion of a SN in 
the ISM is essentially a dynamical process, and therefore requires 
non-stationary calculations.  A general time-dependent model for the 
interaction of EPs in the ISM has been developed and presented in 
Parizot (1999), so we shall use it here extensively, recalling only 
the results relevant to our specific problem and calculating the 
required inputs for processes~1 and~2.

\subsection{The mathematical formalism and the physical ingredients}

In each case, we separate the acceleration of the energetic particles 
(EPs) from their propagation and interaction within the SNR. This is 
legitimate because the time scale for acceleration up to the energies 
we are concerned with is very much smaller than any other time scale 
in the problem, whether dynamical (SNR evolution) or physical (energy 
loss rate, spallation rates).  Consequently, our calculations apply to 
the EPs once they have been `injected' inside the SNR (from the region 
close to the shock).  Let us assume for the moment that we have 
determined the so-called \textit{injection function}, $Q_{i}(E,t)$, 
which we define as the number of particles of species $i$ introduced 
at energy $E$ and time $t$, per unit energy and time (in 
$(\mathrm{MeV/n})^{-1}\mathrm{s}^{-1}$).  The EP distribution 
function, $N_{i}(E,t)$ then satisfies the usual propagation equation 
(see Parizot 1999)~:

\begin{equation}
\begin{split}
	\frac{\partial}{\partial t}N_{i}(E,t) &+ \frac{\partial}
	{\partial E}(\dot E_{i}(E) N_{i}(E,t) )\\
	&= Q_{i}(E,t) + Q_{i}^{\prime}(E,t) -
	\frac{N_{i}(E,t)}{\tau_{i}^{\mathrm{tot}}(E)},
	\label{PropEqOneZoneOnePhase}
\end{split}
\end{equation}
where $\dot E_{i}(E)$ is the energy loss rate for the nuclei of 
species $i$ at energy $E$ (in $(\mathrm{MeV/n})\mathrm{s}^{-1}$), 
$Q_{i}^{\prime}(E,t)$ is the production rate of nuclei $i$ as 
secondary particles, and $\tau_{i}^{\mathrm{tot}}(E)$ is the time 
scale for catastrophic losses, such as nuclear destruction or escape 
from the region under study.

Since we are concerned with spallation reactions involving the nuclei 
of the CNO group, we can neglect the two-step processes such as 
$^{16}\mathrm{O} + \mathrm{p} \rightarrow$ $^{12}\mathrm{C}$ followed 
by $^{12}\mathrm{C} + \mathrm{p} \rightarrow$ $^{9}\mathrm{Be}$.
We indeed found, using a steady-state model, that the omission of the 
two-step processes leads too an error of at most $\sim 10\%$, in
good agreement with Ramaty et al.  (1997) calculations.  Since this is 
smaller than the other observational and theoretical uncertainties, 
and their implementation in a time-dependent model greatly complicates 
the situation, we shall neglect them here (Note that in any case, even 
if there were no other uncertainty in the problem, it is much more 
accurate to do time-dependent calculations without two-step processes 
than steady-state calculations including two-step processes, as our 
simulations have shown).  To state this in a more physical way, we can 
claim that the spallative production of carbon amounts to at most a 
few percents of the initial CNO supply from the supernova explosion.  
To the level of precision of the SN models, to mention that only, this 
correction is of no significance, so we shall simply drop 
$Q_{i}^{\prime}(E,t)$ in Eq.~(\ref{PropEqOneZoneOnePhase}).

Concerning the catastrophic loss time, $\tau_{i}^{\mathrm{tot}}$, it 
is obtained for stable nuclei as~:
\begin{equation}
	\frac{1}{\tau_{i}^{\mathrm{tot}}(E,t)} = 
	\frac{1}{\tau_{i}^{\mathrm{esc}}(E,t)} +
	\frac{1}{\tau_{i}^{\mathrm{D}}(E,t)},
	\label{tauTot}
\end{equation}
where $\tau_{i}^{\mathrm{esc}}$ is the escape time, and 
$\tau_{i}^{\mathrm{D}}$ is the destruction time.  The latter is 
derived from semi-empirical formulas giving the total 
inelastic cross sections $\sigma_{i,j}$ for a projectile $i$ in a 
target of species $j$ (Silberberg \& Tsao 1990), according to~:
\begin{equation}
	\frac{1}{\tau_{i}^{\mathrm{D}}(E,t)} =
	[\sum_{j}\sigma_{i,j}(E)n_{j}(t)]v(E),
	\label{destructionTime}
\end{equation}
where $v(E)$ is the velocity of the energetic particle and $n_{j}(t)$ 
is the number density of target species $j$ at time $t$.

Following the above qualitative analysis (see Paper~I for more 
details), we assume that the time of escape out of the SNR is infinite 
during the Sedov-like phase of the SNR expansion, and `zero' afterwards.  
This merely translates the fact that the EPs are confined within the 
SNR during the adiabatic phase (at least those of lowest energy, which 
produce most of the spallative LiBeB), and then leak out on a very 
short time scale.  Once the EPs have escaped from the SNR, we need to 
distinguish between our two processes.  In the first case 
(acceleration at the forward shock), the EPs are deprived of CNO and 
will not give rise to enough spallation reactions out of the SNR to 
raise the LiBeB production in any significant way.  This is due to the 
very low ambient metallicity.  In the second case, however, the EPs 
are made of the supernova ejecta themselves and are thus rich in CNO. 
As a consequence, as far as LiBeB production is concerned, there is no 
difference whether they interact within or outside the SNR, as 
interactions with H and He nuclei dominate anyway.  We must therefore 
follow these accelerated nuclei after the end of the adiabatic phase, 
and compute the corresponding contribution to the total production of 
light elements.

Concerning the energy loss rate, $\dot E_{i}(E)$, we need to take into 
account both ionisation (Coulombian) and adiabatic losses.  The former 
are very common and just cannot be avoided as soon as energetic 
particles are to be interacting in the ISM. The latter, however, must 
be included here because the EPs are confined within the SNR where 
their velocities are randomized.  As a consequence, they do 
participate to the internal pressure which drives the remnant during 
the Sedov-like phase, and suffer the adiabatic losses like any other 
particle working outward when reflected at the expanding shell.  
Quantitatively, these adiabatic losses have been calculated in 
Paper~I. They are given by Eq.~(14) there, namely~:
\begin{equation}
	\frac{\dot{p}}{p} = -\frac{3}{4}\frac{\dot{R}}{R}\,\,,
	\label{AdiabLossRate}
\end{equation}
where $p$ is the momentum of the particle and $R(t)$ is the radius of 
the shock.  Assuming the Sedov-like expansion law ($R(t)\propto 
t^{2/5}$) and writing the loss rate in terms of energy, we obtain~:
\begin{equation}
	\dot E_{\mathrm{ad}}(E,t) = -\frac{3}{10}\frac{E}{t}
	\left(\frac{E + 2m_{p}c^{2}}{E + m_{p}c^{2}}\right)
	\label{adiabaticLosses2}
\end{equation}

This energy loss rate does not depend on the EP species, but is 
clearly a function of time.  On the other hand, the ionisation 
losses, $\dot E_{\mathrm{ion}}(E)$, do depend on the nuclear 
species, as well as on time, indirectly, through the density and 
composition of the ambient medium.  Indeed, it has to be realised 
that the medium in which the EPs are `propagating', namely the 
interior of the SNR, is initially very rich in freshly synthesized 
CNO nuclei, and then gets poorer and poorer in metals as the 
ejecta are being diluted in the ambient, metal-poor, swept up 
material.

This \textit{dilution effect} is most important for the calculation of 
the total LiBeB production through our first mechanism (acceleration 
of the ISM at the forward shock).  Indeed, the instantaneous 
production rates are directly proportional to the density of CNO 
within the remnant at time $t$, which goes like $R^{-3}$, i.e.  
$t^{-6/5}$.  Quantitatively, the LiBeB production rates are obtained 
by integrating the spallation cross sections over the EP distribution 
functions~:
\begin{equation}
	\frac{\d N_{k}}{\d t} =
	\sum_{i,j}\int_{0}^{\infty}\mathrm{d}E^{\prime}
	N_{i}(E^{\prime},t) n_{j}(t)
	\sigma_{i,j;k}(E^{\prime}) v_{i}(E^{\prime}),
	\label{ProductionRates}
\end{equation}
where $\sigma_{i,j;k}$ is the cross section for the reaction $i + j 
\rightarrow k$, and $n_{j}$ is the number density of nuclei $j$ in the 
target (here, the interior of the SNR).

The total LiBeB production is then obtained for the first mechanism by 
integrating these production rates from $\tSW$ to $\tend$, which marks 
the end of the Sedov-like phase as well as the end of the confinement 
of the EPs inside the SNR. For the second mechanism, we need to 
integrate from $t_{\mathrm{SW}}$ to the confinement time of the cosmic 
rays within the Galaxy.  As we shall see below, integrating up to 
infinity only leads to a small overestimate of the total LiBeB 
production, since the low energy cosmic rays responsible for most of 
that production have anyway a short lifetime above the spallation 
thresholds.

The sweep-up time, \tSW, is obtained straightforwardly from its 
definition as a function of the SN parameters and the ambient number 
density, $n_{0}$~:
\begin{equation}
	\tSW = (1.4\,10^{3}\,\mathrm{yr})
	\left(\frac{M_{\mathrm{ej}}}{10\Msun}\right)^{\hspace{-2pt}\frac{5}{6}}
	\hspace{-4pt}
	\left(\frac{E_{\mathrm{SN}}}{10^{51}\mathrm{erg}}\right)^{\hspace{-2pt}-\frac{1}{2}}
	\hspace{-4pt}
	\left(\frac{n_{0}}{1\mathrm{cm}^{-3}}\right)^{\hspace{-2pt}-\frac{1}{3}}.
	\label{SweepUpTime}
\end{equation}
The determination of \tend~is more difficult and somewhat arbitrary, 
even in the approximation of a perfectly homogeneous circumstellar 
medium.  We argued above and in Paper~1 that \tend~should more or less 
coincide with the end of the Sedov-like phase, when the shock induced 
by the SN explosion becomes radiative, that is when the cooling time 
of the post-shock gas becomes of the same order as the dynamical time.  
This depends on the cooling function which in turn depends on the 
density and metallicity of the post-shock gas.  Such details and their 
influence on \tend~have been considered in Paper~I. Here, we only give 
the asymptotic result, valid in the limit of large ambient densities, 
$n_{0}$~:
\begin{equation}
	\tend = (1.1\,10^{5}\,\mathrm{yr})
	\left(\frac{E_{\mathrm{SN}}}{10^{51}\mathrm{erg}}\right)^{1/8}
	\left(\frac{n_{0}}{1\mathrm{cm}^{-3}}\right)^{-3/4}.
	\label{tEnd}
\end{equation}
Comparing the dependence of \tSW~and \tend~on density, we find that 
the Sedov-like phase gets shorter when $n_{0}$ is increased, and thus 
the duration of process~1 decreases.

\subsection{The injection function at the forward shock}

We now turn to the determination of the injection function, 
$Q_{i}(E,t)$, in the case of our first mechanism.  As suggested by 
shock acceleration calculations, we assume that the distribution 
function of the accelerated particles is $f(p) \propto p^{-4}$, so 
that the number of protons injected inside the SNR per unit time 
between momenta $p$ and $p + \d p$, irrespective of their direction, 
is~:
\begin{equation}
	Q(p)\d p = Q_{0}\frac{\d p}{p^{2}},
\end{equation}
from thermal values up to $\sim 10^{14}$~eV/c.  This leaves us only 
with the calculation of the normalisation, $Q_{0}$, as a function of 
time.

Following again the most widely accepted theoretical ideas, we assume 
that the total energy injected per unit time in the form of energetic 
particles at time $t$ is equal to a constant fraction, $\theta_{1}$, 
of the power, $\mathcal{P}_{in}$, flowing through the shock at that 
time (recalling that the acceleration time scale is small as compared 
to the dynamical one).  Mathematically, this normalisation condition 
reads~:
\begin{equation}
	\int_{p_{\mathrm{min}}}^{p_{\mathrm{max}}} Q(p)E(p)\d p
	= \theta_{1}\mathcal{P}_{\mathrm{in}},
	\label{normalisation}
\end{equation}
where $E(p) = \sqrt{p^{2}c^{2} + m^{2}c^{4}} - mc^{2}$ is the energy 
of a proton of impulsion $p$.  Integrating the left hand side (LHS) of 
Eq.~(\ref{normalisation}), one finds~:
\begin{equation}
\begin{split}
	\mathrm{LHS} &=
	Q_{0}c\int_{p_{\mathrm{min}}}^{p_{\mathrm{max}}}\frac{mc}{p}
	\left[\sqrt{1 + (p/mc)^{2}} - 1 \right]\frac{\d p}{p}\\
	&= Q_{0}c\int_{u_{\mathrm{min}}}^{u_{\mathrm{max}}}
	\frac{\sqrt{1 + u^{2}} - 1}{u^{2}}\d u\\
	&\equiv Q_{0}c\kappa,
	\label{variableChange}
\end{split}
\end{equation}
where $u = p/mc$ and $\kappa$ is the number
\begin{equation}
	\kappa = \left[\frac{1 - \sqrt{1 + u^{2}}}{u} +
	\ln(u + \sqrt{1 + u^{2}})
	\right]_{u_{\mathrm{min}}}^{u_{\mathrm{max}}}
	\label{kappa}
\end{equation}
Typical values for $u_{\mathrm{min}}$ and $u_{\mathrm{max}}$ are 
$u_{\mathrm{max}} = p_{\mathrm{max}}/mc \sim 10^{14}/2\,10^{9} \sim 
5\,10^{4}$, and $u_{\mathrm{min}} = 
\sqrt{2E_{\mathrm{min}}/m_{\mathrm{p}}c^{2}} \la 10^{-2}$. Then, to first 
order~:
\begin{equation}
	\kappa = \ln u_{\mathrm{max}} - 1 + \ln 2 + \mathrm{O}(u_{\mathrm{min}}
	+ \mathrm{O}(1/u_{\mathrm{max}}) \sim 10.5
	\label{kappaValue}
\end{equation}
depending on $p_{\mathrm{max}}$ only logarithmically.

Combining Eqs.~(\ref{normalisation}) and~(\ref{variableChange}), we 
obtain the injection function at the forward shock as~:
\begin{equation}
	Q(p) = \theta_{1}\frac{\mathcal{P}_{\mathrm{in}}}{\kappa p^{2}c},
	\label{Q(p)}
\end{equation}
or in terms of energy~:
\begin{equation}
	Q(E) = Q(p)\frac{\d p}{\d E} = 
	\frac{\theta_{1}\mathcal{P}_{\mathrm{in}}}{\kappa}\frac{1}{E^{3/2}}
	\frac{E + mc^{2}}{(E + 2mc^{2})^{3/2}}
	\label{Q(E)}
\end{equation}

The asymptotical behavior is thus~: $Q(E) \propto E^{-1.5}$ for $E\ll 
m_{\mathrm{p}}c^{2}$, and $Q(E) \propto E^{-2}$ for $E\gg 
m_{\mathrm{p}}c^{2}$.

It should be clear that the above injection function is indeed a 
function of time, through the incoming power 
$\mathcal{P}_{\mathrm{in}}$.  To evaluate it, one can make use of the 
well known formulas giving the time evolution of the shock radius, 
$R_{\mathrm{s}}$, and velocity, $V_{\mathrm{s}}$, during the 
Sedov-like phase, and calculate $\mathcal{P}_{in} = 
\frac{1}{2}\rho_{0}V_{\mathrm{s}}^{2}\times 4\pi 
R_{\mathrm{s}}^{2}\times V_{\mathrm{s}}$.  However, since the Sedov 
phase is a similarity solution, we know that the result will be 
nothing else but $\mathcal{P}_{in}(t) \simeq E_{\mathrm{SN}}/t$, where 
$E_{\mathrm{SN}}$ is the explosion energy.  The time-dependent 
injection function is then finally~:

\begin{equation}
	Q(E,t) = \frac{\theta_{1}}{\kappa}\frac{E_{\mathrm{SN}}}{t}
	\frac{1}{E^{3/2}}\frac{E + mc^{2}}{(E + 2mc^{2})^{3/2}}.
	\label{Q(E,t)}
\end{equation}

As can be seen, the power injected in the form of energetic particles 
decreases as $t^{-1}$ as the SNR expands.  This is not a futile 
result, since it happens that the earliest times are also the most 
favourable to the spallative production of light elements in a SNR. 
Indeed, as was discussed in Sect.~\ref{NonStationarity}, the CNO 
nuclei suffer a rapid dilution as the remnant expands, lowering the 
spallation rates.  Ignoring the enhancement of the EPs when the SNR is 
still rather small would thus leads one to significantly underestimate 
the LiBeB production.

In the above derivation, we did not worry about the chemical 
composition of the EPs.  Clearly the injection function still has to 
be weighted by the relative abundance of each nuclear species present 
in the ISM swept up by the SNR. As already mentioned, we are 
interested only in the LiBeB production at low ambient metallicity, 
since this is when the observed proportionality between Be and Fe 
abundances is the most striking and unexpected.  According to the 
assumption that we are testing here, each supernova leads to the same 
amount of $^{9}$Be production, whatever the ambient metallicity.  
Therefore, all our calculations are made with a zero ambient 
metallicity.  The EPs accelerated out of the ISM are thus made of H 
and He only, with their primordial relative abundances.

\subsection{The injection function at the reverse shock}

In the case of the acceleration of the supernova ejecta through the 
reverse shock, the injection function can be written straightforwardly 
as~:
\begin{equation}
	Q_{i}(E,t) = n_{i}\bar{Q}(E)\delta(t - t_{\mathrm{SW}}),
	\label{reverseQ}
\end{equation}
where it is assumed that the acceleration takes place instantaneously 
at $t_{\mathrm{SW}}$.  This may be justified by noting that the 
genuine acceleration and reverse shock evolution time scales are 
certainly smaller than EP evolution time scales (nuclear interactions 
and energy losses). The relative abundance of the different nuclei in 
the accelerated particles just reflects that of the supernova ejecta, 
$n_{i}$, and the shape of $\bar{Q}(E)$ is the same as above.  This 
time, however, the injection function has to be normalised to~:
\begin{equation}
	\sum_{i}\int\d t\int Q_{i}(E,t)E\d E = \theta_{2}E_{\mathrm{SN}},
	\label{normalisation2}
\end{equation}
where $\theta_{2}$ is the fraction of the explosion energy which goes 
into the EPs accelerated at the reverse shock.  This can be 
phenomenologically expressed as the product of two coefficient~: 
$\theta_{2} = \theta_{\mathrm{acc}}\times\theta_{\mathrm{rev}}$, where 
$\theta_{\mathrm{acc}}$ is the fraction of the shock energy imparted 
to the EPs (i.e.  $\theta_{\mathrm{acc}}\approx\theta_{1}$, defined 
above), and $\theta_{\mathrm{rev}}$ is the fraction of the explosion 
energy which goes into the reverse shock.  In our calculation, we 
adopt the `canonical values' of $\theta_{1} = 0.1$ and 
$\theta_{\mathrm{rev}} = 0.1$, and thus $\theta_{2} = 0.01$.  It 
should be clear, however, that these values are only indicative, and 
that the results simply scale proportionally to $\theta_{1}$ and 
$\theta_{2}$.

The time integration in Eq.~(\ref{normalisation2}) is straightforward, 
and with $\sum_{i}n_{i} = 1$, we get~:
\begin{equation}
	\bar{Q}(E) = \frac{\theta_{2}E_{\mathrm{SN}}}{\kappa}
	\frac{1}{E^{3/2}}\frac{E + mc^{2}}{(E + 2mc^{2})^{3/2}},
	\label{Qbar}
\end{equation}
where $\kappa$ has been given in Eq.~(\ref{kappa}) 
and~(\ref{kappaValue}).  Note that the mass $m$ appearing in the above 
expressions is always the proton mass, and that correlatively the 
energies are expressed in MeV/n for all the nuclear species.

\subsection{The formal solution for the EP distribution function}

The formal solution of the time-dependent propagation 
equation~(\ref{PropEqOneZoneOnePhase}) is (Parizot 1999)~:
\begin{equation}
\begin{split}
	N_{i}(E,t) = \frac{1}{|\dot{E}_{i}(E)|}&\int_{E}^{+\infty}
	Q_{i}(E_{0},t - \tau_{i}(E_{0},E))\\
	&\times \exp\Big(-
	\int_{E_{0}}^{E}\frac{\mathrm{d}E^{\prime}}{\dot{E}_{i}(E^{\prime})
	\tau_{\mathrm{tot},i}(E^{\prime})}\Big)\mathrm{d}E_{0},
	\label{FormalSolution}
\end{split}
\end{equation}
where
\begin{equation}
	\tau_{i}(E_{0},E) = \int_{E_{0}}^{E}\frac{\mathrm{d}
	E^{\prime}}{\dot{E_{i}}(E^{\prime})}.
	\label{TauFonction}
\end{equation}

This solution, however, only considers the time-dependence of the 
injection function, $Q(E,t)$, and not that of the conditions of 
propagation, namely the energy losses and the destruction time.  Now 
it is clear that the adiabatic losses do depend on time as well as the 
ionisation losses and the nuclear destruction time, through the 
chemical composition within the SNR. One then needs to divide the 
whole process into sufficiently short phases so that these parameters 
stay approximately constant during each phase, and put together the 
solutions (\ref{FormalSolution}) for each phase in a proper way (for 
details, see Parizot, 1999).  For the present calculations, it proved 
sufficient to divide the Sedov-like phase into 15 successive phases.

In the case of our second injection function, Eq.~(\ref{reverseQ}), 
corresponding to the reverse shock acceleration, the time delta 
function allows us to integrate Eq.~(\ref{FormalSolution}) to obtain~:
\begin{multline}
	N_{i}(E,t) = \frac{|\dot{E}_{i}(E_{\mathrm{in}})|}{|\dot{E}_{i}(E)|}
	n_{i}\bar{Q}(E_{\mathrm{in}})\\
	\exp\Big(-\int_{E_{\mathrm{in}}}^{E}\frac{\mathrm{d}E^{\prime}}
	{\dot{E}_{i}(E^{\prime})\tau_{\mathrm{tot},i}(E^{\prime})}\Big),
	\label{IntegratedSolution}
\end{multline}
where $E_{\mathrm{in}}(i,E,t)$ is the solution of~:
\begin{equation}
	\int_{E_{\mathrm{in}}}^{E}\frac{\d 
	E^{\prime}}{\dot{E}_{i}(E^{\prime})} = t - t_{\mathrm{SW}}.
	\label{E_in}
\end{equation}
In other words, $E_{\mathrm{in}}$ is the energy at which a particle of 
species $i$ must have been accelerated at time $t_{\mathrm{SW}}$ in 
order to have slowed down to energy $E$ at time $t$.  Similarly, the 
exponential factor in Eqs~(\ref{FormalSolution}) 
and~(\ref{IntegratedSolution}) is nothing but the survival probability 
of a particle $i$ from its injection at energy $E_{0}$ (or 
$E_{\mathrm{in}}$) to the current energy, $E$.

The above solutions allow us to calculate the EP distribution function 
for both of our injection functions, Eqs~(\ref{Q(E,t)}) 
and~(\ref{reverseQ})--(\ref{Qbar}).  Equation~(\ref{ProductionRates}) 
can then be used to compute the LiBeB production rates at any time 
after the beginning of acceleration, at $t_{\mathrm{SW}}$.  The 
results are presented in the following section.

\section{The results}
\label{Results}

\subsection{LiBeB production by the EPs from the forward shock}

The results we show in this section are obtained with the SN explosion 
models calculated by Woosley and Weaver (1995).  We use their models 
Z, U and T, corresponding to stars with initial metallicity $Z = 0$, 
$10^{-4}\,Z_{\odot}$, and $10^{-2}\,Z_{\odot}$, respectively, and keep 
the same labels as the authors to refer to specific models (e.g.  
model U15A corresponds to a star of 15~\Msun~with 
$10^{-4}\,Z_{\odot}$ initial metallicity and a standard explosion 
energy of $\approx 1.2\,10^{51}$~erg).  We adopt the value $\theta_{1} 
= 0.1$ throughout, on the understanding that all the spallation rates 
are merely proportional to this parameter.

\begin{figure}
\resizebox{\hsize}{!}{\includegraphics{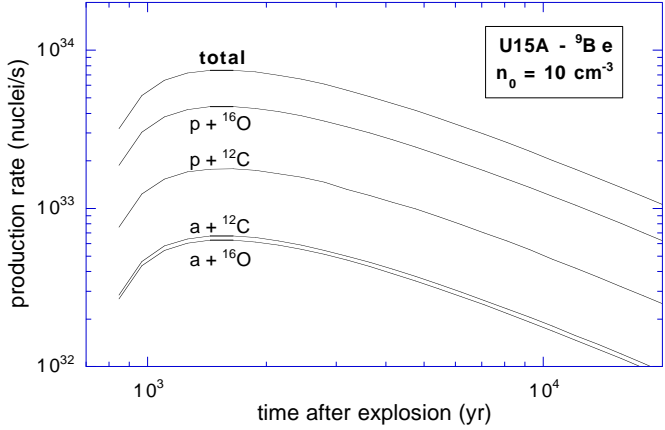}} 
\caption{Process~1 $^{9}$Be production rate in numbers of nuclei per second 
through different spallation reactions as a function of time after the 
SN explosion.  The SN model used is U15A (from Woosley and Weaver, 
1995), and the ambient density is $n_{0} = 10\,\mathrm{cm}^{-3}$.}
\label{SNR2_f1}
\end{figure}

In Fig.~\ref{SNR2_f1}, we show the typical evolution of the 
spallation rates for Be production as a function of time, for a SN 
exploding in a medium with mean density $n_{0} = 
10\,\mathrm{cm}^{-3}$.  The main contribution is seen to come from 
reaction $\mathrm{p} + ^{16}$O, which is due to the low 
$\mathrm{C}/\mathrm{O}$ abundance ratio in the SN ejecta.  For 
reactions involving alpha particles, this deficiency of carbon as 
compared to oxygen is compensated by a greater spallation efficiency.  
The general shape of the curves is easily understood if one refers to 
Eq.~(\ref{ProductionRates}) and to the analysis of the preceding 
section.  Indeed, the spallation rates are basically the product of 
the relevant cross section by the spectral density of energetic 
protons, $N_{\mathrm{p}}(t)$, and the number density of Oxygen within 
the SNR. Now the latter is subject to dilution by the swept-up 
metal-free gas, and therefore decreases as $R^{-3}$, or 
$R_{\mathrm{SW}}^{-3}(t/\tSW)^{-6/5}$, while $N_{\mathrm{p}}(t)$ is 
merely the time integral of the injection function, Eq.~(\ref{Q(E,t)}) 
(at least as long as one can neglect the energy losses).  We thus find 
$N_{\mathrm{p}}(t) \propto \ln(t/\tSW)$, and the spallation rates~:

\begin{equation}
	\frac{\d N_{\mathrm{Be}}}{\d t} \propto R_{\mathrm{SW}}^{-3}
	(t/\tSW)^{-6/5}\ln\left(\frac{t}{\tSW}\right),
	\label{tMax}
\end{equation}
which fits very well the curves in Fig.~\ref{SNR2_f1}.  
Differentiating the above expression, we find the maximum production 
rates to occur at $t = e^{5/6}\tSW \approx 2.3\,\tSW$, which expresses 
the best compromise between Oxygen dilution in the SNR and a 
sufficient injection of EPs since the onset of the acceleration 
process.

\begin{figure}
\resizebox{\hsize}{!}{\includegraphics{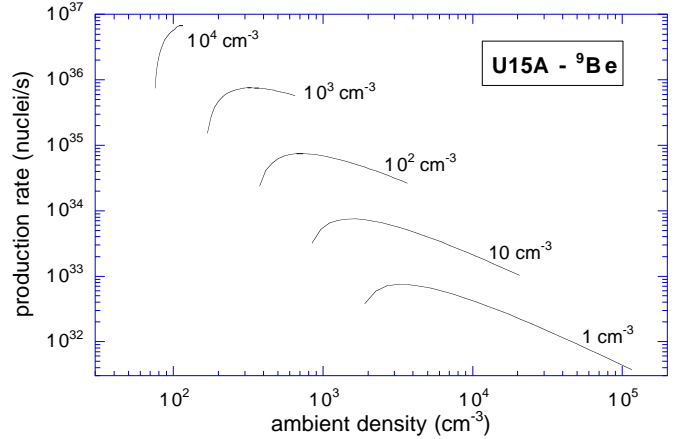}} 
\caption{Process~1 total $^{9}$Be production rates as a function of time after 
the explosion of SN model U15A, for different ambient densities. Each 
curve starts shortly after the sweep-up time and ends at the 
adiabatic time, marking the end of the Sedov-like phase.}
\label{SNR2_f2}
\end{figure}

This behavior can be further observed on Fig.~\ref{SNR2_f2} where 
we plot the total production rates of Be as a function of time after 
explosion, for different values of the ambient density, ranging from 1 
to $10^{4}\,\mathrm{cm}^{-3}$.  The shortening of the Sedov-like phase 
already mentioned is clearly apparent on the figure, as is the 
behavior of $\tSW \propto n_{0}^{-1/3}$ and $\tend \propto 
n_{0}^{-3/4}$.  The calculations also confirm that the position of the 
maximum is always at $t_{\mathrm{max}} \approx 2.3\,\tSW$, although at 
the highest densities, this is very close indeed to the end of the 
adiabatic expansion phase, when the confinement of the EPs ceases and 
the whole process stops.  The position of the maximum then varies as 
$n_{0}^{-1/3}$, while its height, obtained by replacing $t$ by 
$t_{\mathrm{nax}}$ in Eq.~(\ref{tMax}), is proportional to 
$R_{\mathrm{SW}}^{-3}$, and thus $n_{0}$.

\begin{figure}
\resizebox{\hsize}{!}{\includegraphics{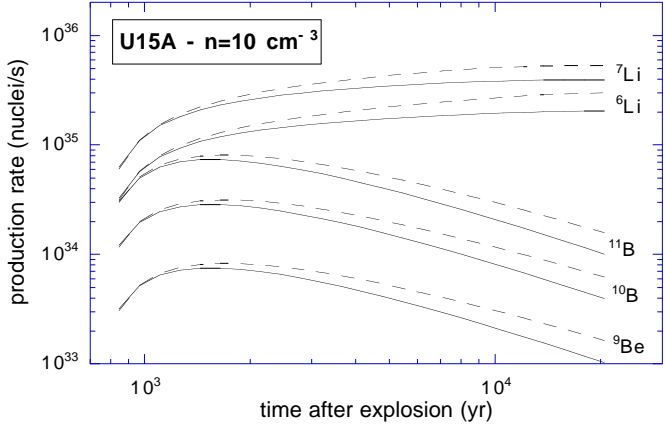}} 
\caption{Process~1 production rates of the five light element isotopes as a 
function of time after the explosion of SN model U15A, in a medium of 
density $n_{0} = 10\,\mathrm{cm}^{-3}$.  Results are shown for 
calculations taking adiabatic losses into account (full lines) as well 
as ignoring them (dashed lines).}
\label{SNR2_f3}
\end{figure}

\begin{figure}
\resizebox{\hsize}{!}{\includegraphics{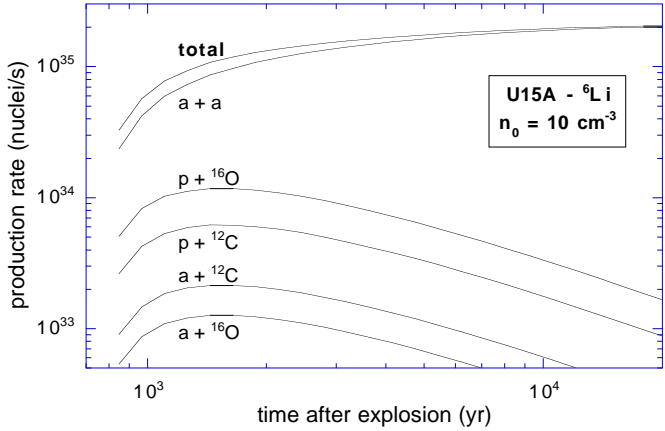}} 
\caption{Process~1 $^{6}$Li production rate in numbers of nuclei per second 
through different spallation reactions as a function of time after the 
SN explosion.  The SN model used is U15A and the ambient density is 
$n_{0} = 10\,\mathrm{cm}^{-3}$.}
\label{SNR2_f4}
\end{figure}

In Fig.~\ref{SNR2_f3} we show the evolution of the production 
rates for the five light element isotopes, either taking and not 
taking the adiabatic losses into account.  The behavior of $^{6}$Li 
and $^{7}$Li is different from that of the other isotopes, because 
lithium is mainly produced through $\alpha + \alpha$ reactions, as 
shown in Fig.~\ref{SNR2_f4}, and these reactions are not sensitive to 
the dilution of the SN ejecta by the ambient material.  The evolution 
of Li production rates therefore reflects directly the evolution of 
the EP fluxes.  As just stated, this would be a pure logarithm if one 
could neglect the energy losses.  It turns out that the adiabatic 
losses dominate the Coulombian losses for any reasonable ambient 
density.  To see how they influence the EP fluxes, let us re-write 
Eq.~(\ref{PropEqOneZoneOnePhase}) in the form~:

\begin{equation}
	\frac{\partial}{\partial t}N(E,t) = Q(E,t) - \frac{\partial}
	{\partial E}(\dot E_{\mathrm{ad}}(E,t) N(E,t)),
	\label{SimplifiedPropEq}
\end{equation}
where we dropped the destruction and second order terms. At energies 
of a few tens of MeV/n, Eq.~(\ref{adiabaticLosses2}) simplifies to 
give the expression for adiabatic losses~:

\begin{equation}
	\dot E_{\mathrm{ad}}(E,t) = -\frac{6}{10}\frac{E}{t}
\end{equation}

Replacing in Eq.~(\ref{SimplifiedPropEq}), we obtain~:

\begin{equation}
\begin{split}
	\frac{\partial}{\partial t}N &= Q + 
	\frac{6}{10\,t}\frac{\partial}{\partial E}(EN)\\
	&= Q - \frac{6(\alpha - 1)}{10\,t}N,
\end{split}
\end{equation}
where we recognized that a power-law for the injection function $Q$, 
with spectral index $-\alpha$ ($Q = Q_{0}E^{-\alpha}/t$), translates 
into a power-law for the EP spectral density $N$ with the same index~: 
$N = N_{0}(t)E^{-\alpha}$.  This is a consequence of the 
proportionality between the energy loss rate and the energy itself.  
The equation for $N_{0}$ is then straightforward~:

\begin{equation}
	\frac{\partial}{\partial t}N_{0} = \frac{Q_{0}}{t} - \frac{6(\alpha - 
	1)}{10\,t}N_{0},
	\label{EqForN0}
\end{equation}
from where we see that instead of the logaritmic increase $N(E,t) = 
Q_{0}E^{-\alpha}\ln(t/\tSW)$ prevailing in the absence of energy 
losses, a steady-state value should be reached (if the Sedov-like 
phase last long enough) with~:
\begin{equation}
	N_{0} = \frac{10Q_{0}}{6(\alpha - 1)}.
\end{equation}

So the adiabatic losses are important when both terms in the right 
hand side of Eq.~(\ref{EqForN0}) are of the same order, that is
(evaluating the second term from its `no-loss value', and using 
$\alpha = 1.5$ for the low-energy part of the spectrum)~:
\begin{equation}
	\ln\left(\frac{t}{\tSW}\right) \approx \frac{6}{10}(\alpha - 1),
\end{equation}
or
\begin{equation}
	t\la \tSW\,e^{\frac{6}{10}(\alpha - 1)}\approx 1.35\,\tSW.
\end{equation}

This result is in very good agreement with the numerical results shown 
in Fig.~\ref{SNR2_f3}.  Likewise, the gap between the calculations 
with adiabatic losses turned on or off is increasing only 
logarithmically with time, so that the difference is rather small, 
even at the end of the Sedov-like phase.  We find total Be production 
only a few tens of percent higher if we drop the adiabatic losses, and 
the difference even falls to zero when higher ambient densities are 
considered.  This is of course because the Sedov phase is then 
considerably shortened.

\begin{figure}
\resizebox{\hsize}{!}{\includegraphics{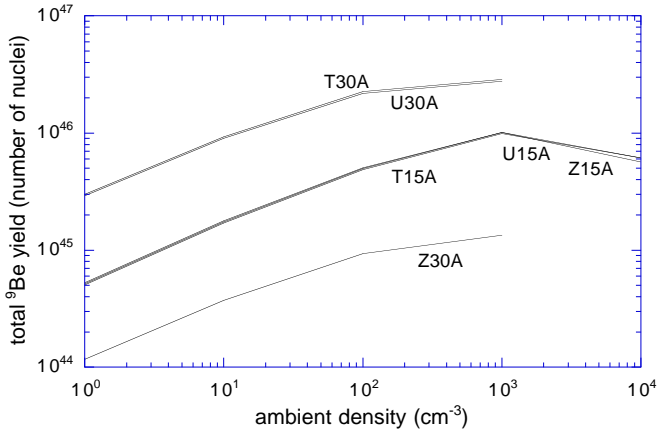}} 
\caption{Integrated process~1 Be yields for different SN models as a 
function of ambient density.  For the models with progenitor masses of 
30~\Msun~and a density higher than a few $10^{3}\,\mathrm{cm}^{-3}$, 
the sweep-up time $\tSW$ is greater than the adiabatic time $\tend$, 
so that the Sedov-like phase does not exists.}
\label{SNR2_f5}
\end{figure}

Although Fig.~\ref{SNR2_f1}, \ref{SNR2_f2} and \ref{SNR2_f3} help us 
to clarify the dynamics of the process and understand the role of the 
different parameters, only the total, integrated light elements 
production is actually relevant to the Galactic chemical evolution.  
We show in Fig.~\ref{SNR2_f5} the results of the integration of the Be 
production rates over the whole Sedov-like phase, for different SN 
explosion models, as a function of the ambient density.  Except for 
the case of the Z30A model, we find that for a given mass of the 
progenitor the total Be yield is independent of the initial 
metallicity of the star (zero, $10^{-4}$ or $10^{-2}$ times solar).  
The very small production of Be obtained with the Z30A model is in 
fact due to a very small amount of Oxygen expelled by the supernova.  
A model with a higher explosion energy (Z30B) gives results closer to 
those of T30A and U30A. Although yields significantly different are 
obtained for different masses of the progenitor, due to different 
compositions and masses of the ejecta, it is clear from 
Fig.~\ref{SNR2_f5} that the total amount of Be produced by process~1 
(forward shock) is much too low to account for the Be observed in 
metal-poor star.  Indeed, the results obtained for a $15\,\Msun$ star 
with ambient density $n_{0} = 1\,\mathrm{cm}^{-3}$ are about three 
orders of magnitude too low, for our choice of $\theta_{1} = 0.1$.  
This is in very good agreement with the analytical estimates presented 
in Paper~I.

Concerning the density dependence of the Be yields, the numerical 
results shown in Fig.~\ref{SNR2_f2} are also in good agreement with 
the analytical calculations.  In particular, the yields increase with 
ambient density and reach a maximum at about a few 
$10^{3}\,\mathrm{cm}^{-3}$, above which the Sedov-like phase becomes 
extremely short, and even vanishes for high mass progenitors (implying 
large ejected masses).  Using Eq.~(\ref{SweepUpTime}) and~(\ref{tEnd}) 
we can write this limiting density as~:
\begin{equation}
	n_{\mathrm{lim}} \simeq (4\,10^{4}\,\mathrm{cm}^{-3})
	\left(\frac{M_{\mathrm{ej}}}{10\,\Msun}\right)^{-2}
	\left(\frac{E_{\mathrm{SN}}}{10^{51}\,\mathrm{erg}}\right)^{3/2}.
	\label{nLim}
\end{equation}

\subsection{LiBeB production by the EPs from the reverse shock}

\begin{figure*}
\centerline{\psfig{file=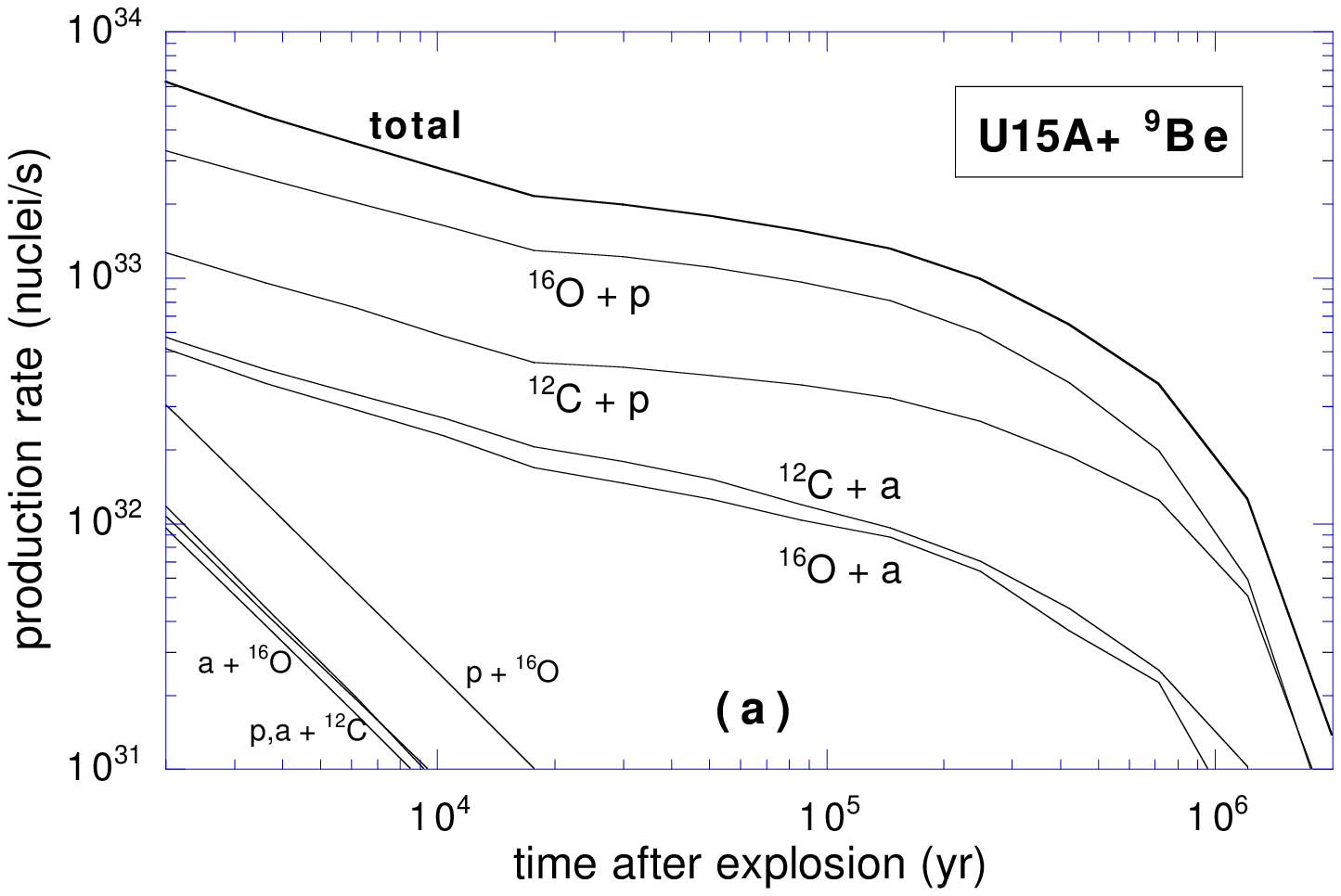, width=8.5cm}
            \hfill
            \psfig{file=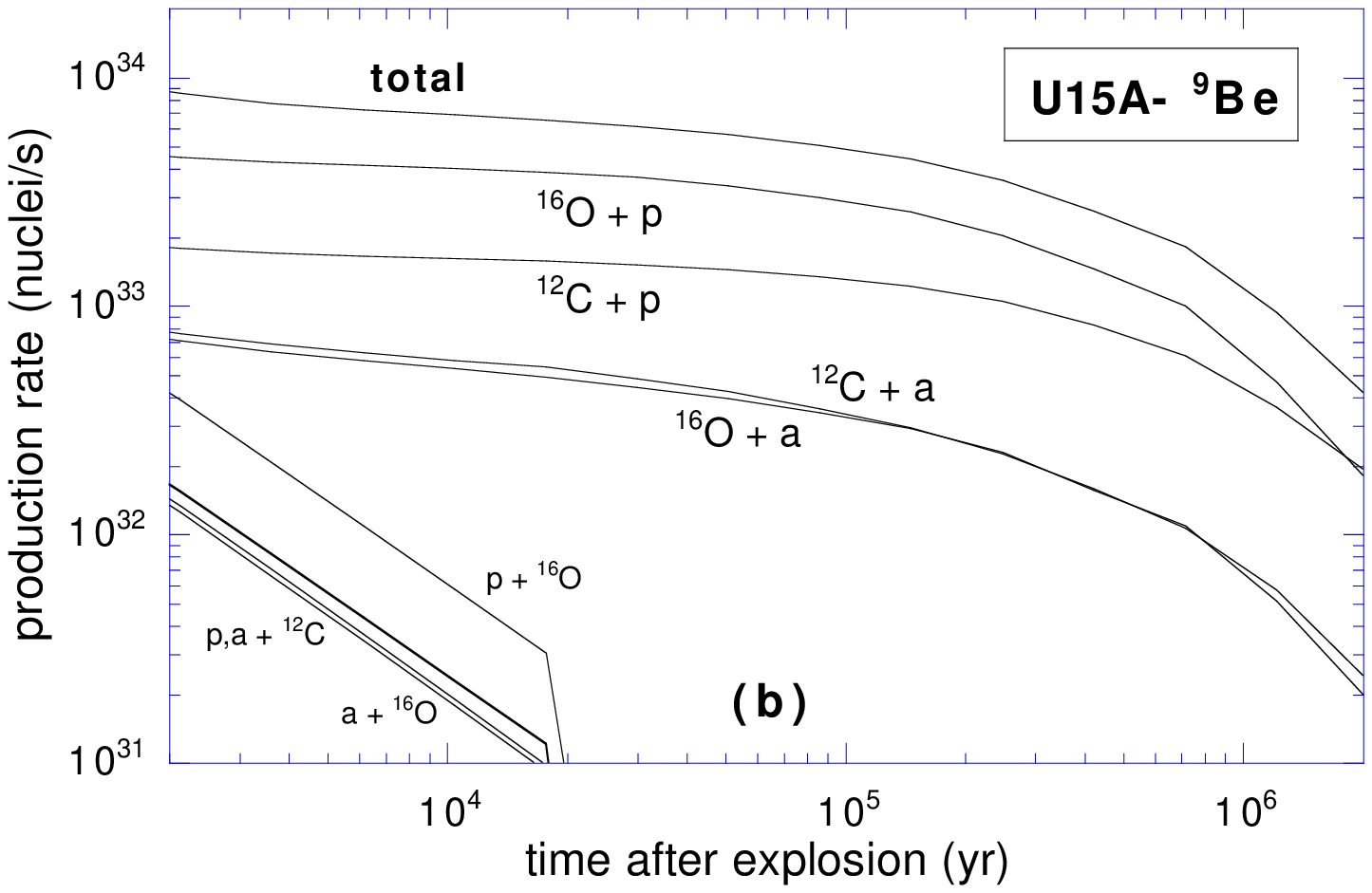, width=8.5cm}}
\centerline{\psfig{file=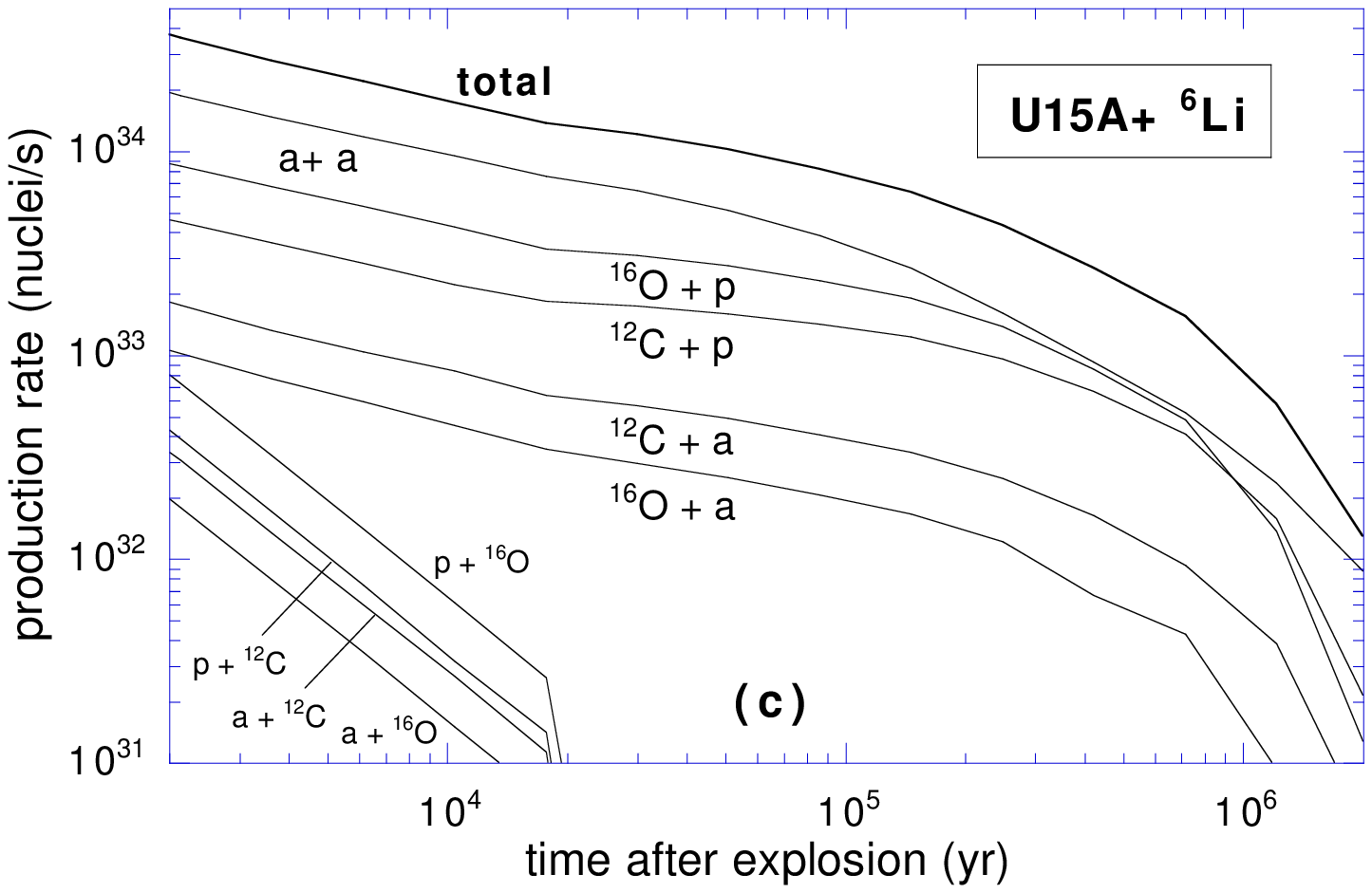, width=8.5cm}
            \hfill
            \psfig{file=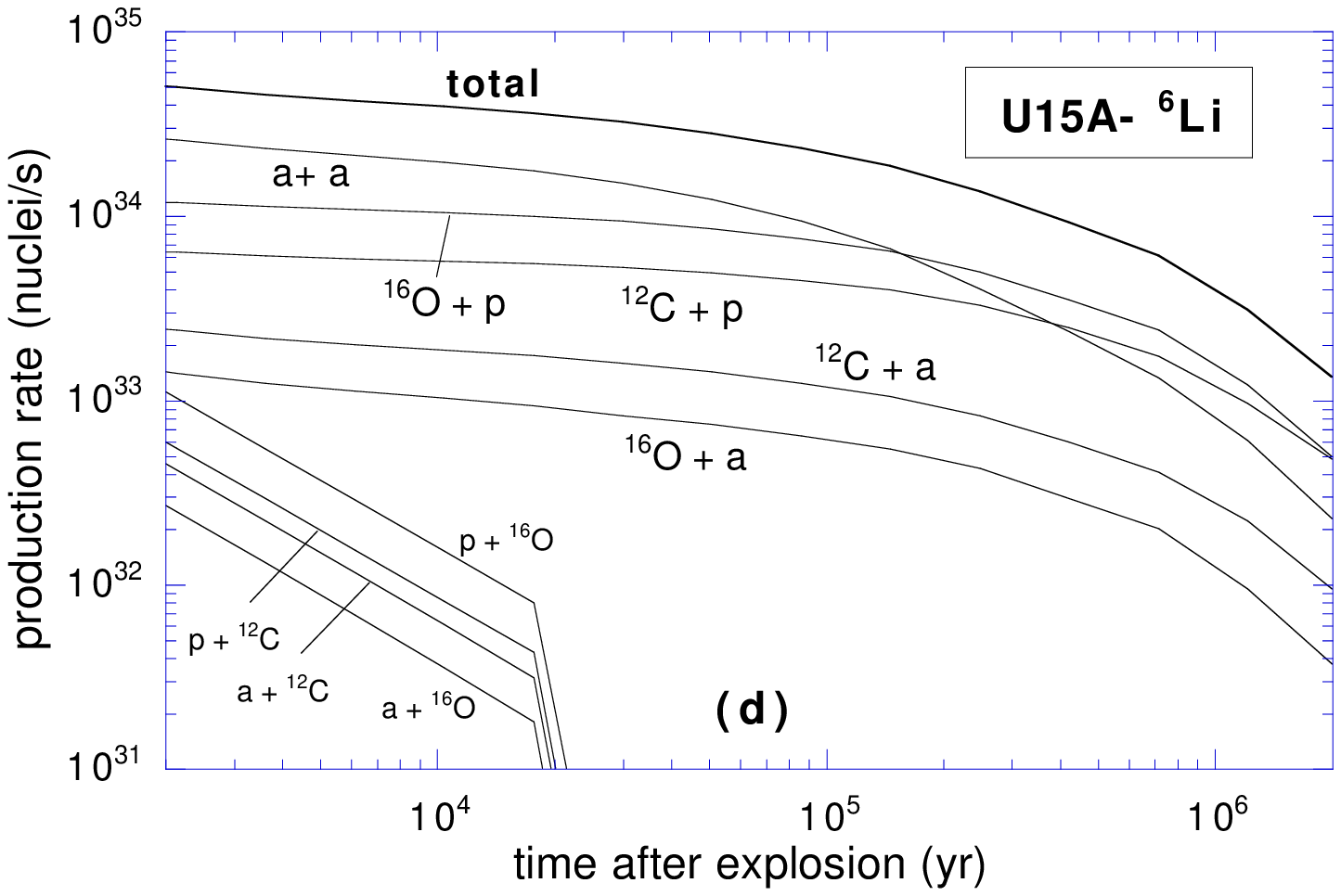, width=8.5cm}}
\caption{Process~2 (and 3) $^{6}$Li and $^{9}$Be production rates in 
numbers of nuclei per second through different spallation reactions as 
a function of time after the SN explosion.  The SN model used is U15A 
and the ambient density is $n_{0} = 10\,\mathrm{cm}^{-3}$.  On the 
left, the calculations include adiabatic losses (U15A+); on the right, 
they do not (U15A-).}
\label{SNR2_f6}
\end{figure*}

We now turn to the results obtained for the second mechanism, in which 
the SN ejecta are accelerated at the reverse shock at the onset of the 
Sedov-like phase.  The $^{6}$Li and $^{9}$Be production rates are 
shown on Fig.~\ref{SNR2_f6} as a function of time, with and without 
adiabatic losses, for an ambient density of $10\,\mathrm{cm}^{-3}$ and 
a progenitor corresponding to the U15A model of SN. As can be seen, 
the Be production rates are strongly dominated by inverse spallation 
reactions, i.e.  reactions in which the projectile is the heavier 
nuclei.  Moreover, since the abundance of C and O in the target 
suffers from dilution by ISM gas, the direct-to-inverse spallation 
efficiency ratio keeps decreasing during the Sedov-like phase.  At the 
end of it, as already discussed, the direct reactions stop, while 
inverse ones are not affected.  In Figs.~\ref{SNR2_f6}b 
and~~\ref{SNR2_f6}d, the adiabatic losses have not been taken into 
account.  The decrease of the direct spallation rates is thus due only 
to dilution, and we obtain the expected power law in $R^{-3}$, or 
$t^{-6/5}$.  In the meanwhile, the inverse spallation rates are almost 
constant, as the Sedov-like phase is much shorter than the time-scale 
for coulombian losses.  This time-scale can literally be read from the 
figure.  It is of order a few times $10^{5}$ years for this set of 
parameters.  Note however that the energy loss time-scale actually 
depends on the species and energy of the particle.  Accordingly, what 
is observed on the spallation rates is in fact a mean coulombian 
time-scale, averaged over the EP energy spectrum, and more precisely 
the part of this spectrum which stands above the energy threshold of 
the cross-sections.  This explains the slight variation observed for 
the different spallation channels.

It is worth emphasizing that the time-scales that we obtain are much 
shorter than the confinement time-scales inferred from cosmic-ray 
propagation theories.  This indicates that the leakage of the EPs out 
of the Galaxy has negligible influence on the spallation yields, and 
justifies our choice of neglecting it.  Even for an ambient density of 
$1\,\mathrm{cm}^{-3}$, the bulk of the light element production is 
contributed by nuclear reactions occuring within a few million years 
after the SN explosion, which is to be compared with Galactic 
confinement times of order a few $10^{7}$~years.

Comparing Fig.~\ref{SNR2_f6}a with Fig.~\ref{SNR2_f6}b (or 
Fig.~\ref{SNR2_f6}c with Fig.~\ref{SNR2_f6}d), we can see the 
influence of the adiabatic losses on the nuclear rates.  For inverse 
spallation reactions, we observe an almost perfect power law decrease, 
with logarithmic slope $\sim 0.4$, in very good agreement with the 
value derived in Paper~I. Indeed, the analytic treatment led us to 
expect spallation rates proportional to $R^{-3/4}$, or equivalently 
$t^{-3/10}$.  The slightly quicker decrease found in the numerical 
results is due to the contribution of the coulombian losses (whose 
effect is also visible on Figs.~\ref{SNR2_f6}b and~\ref{SNR2_f6}d), 
and to the shape of the spallation cross-sections close to their 
threshold.  Likewise, the time evolution of direct spallation 
reactions is also very close to a power law, with logarithmic slope of 
$\sim 1.6 \sim (1.2 + 0.3)$, as expected.

\begin{figure*}
\centerline{\psfig{file=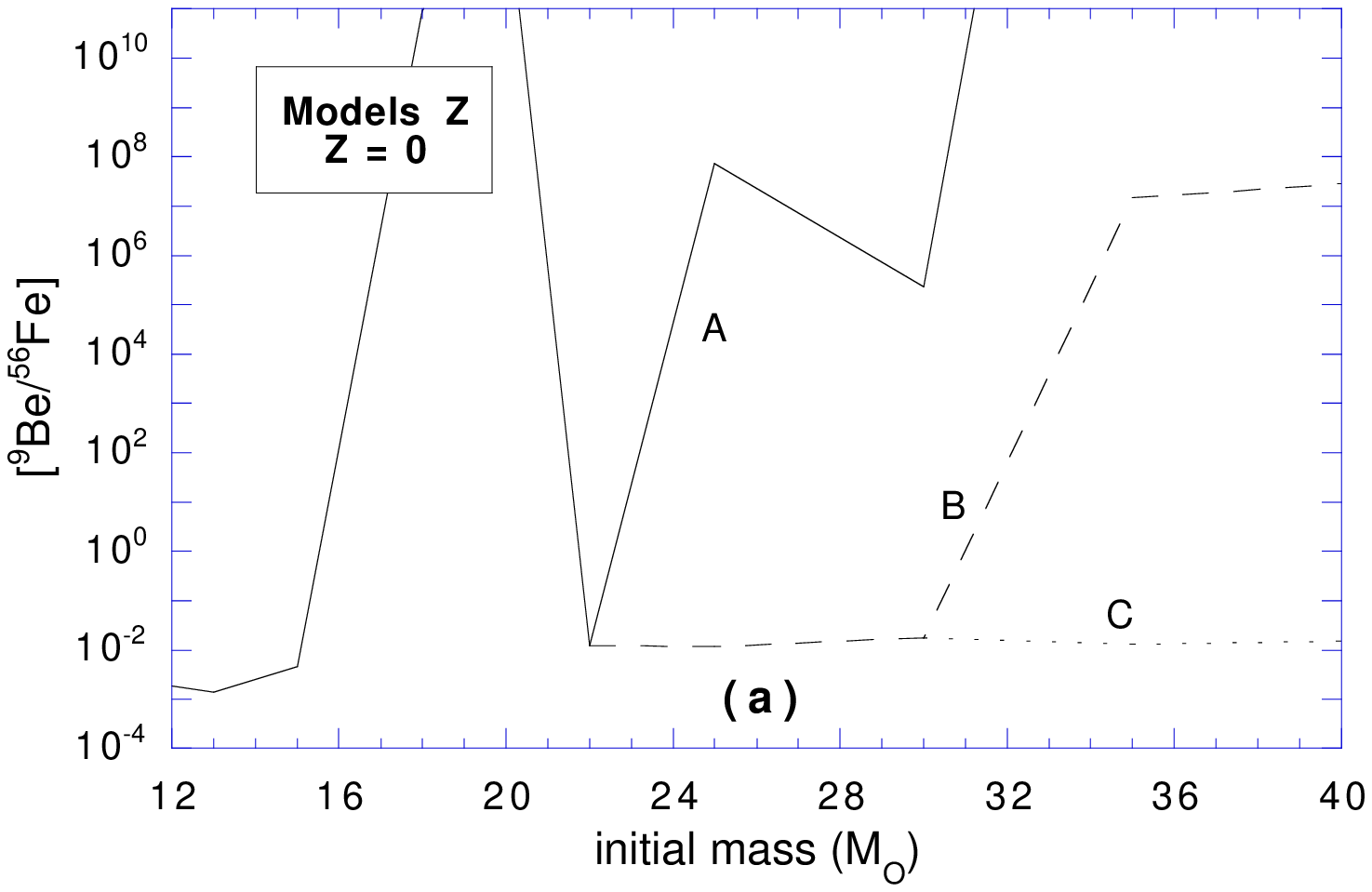, width=8.8cm}
            \hfill
            \psfig{file=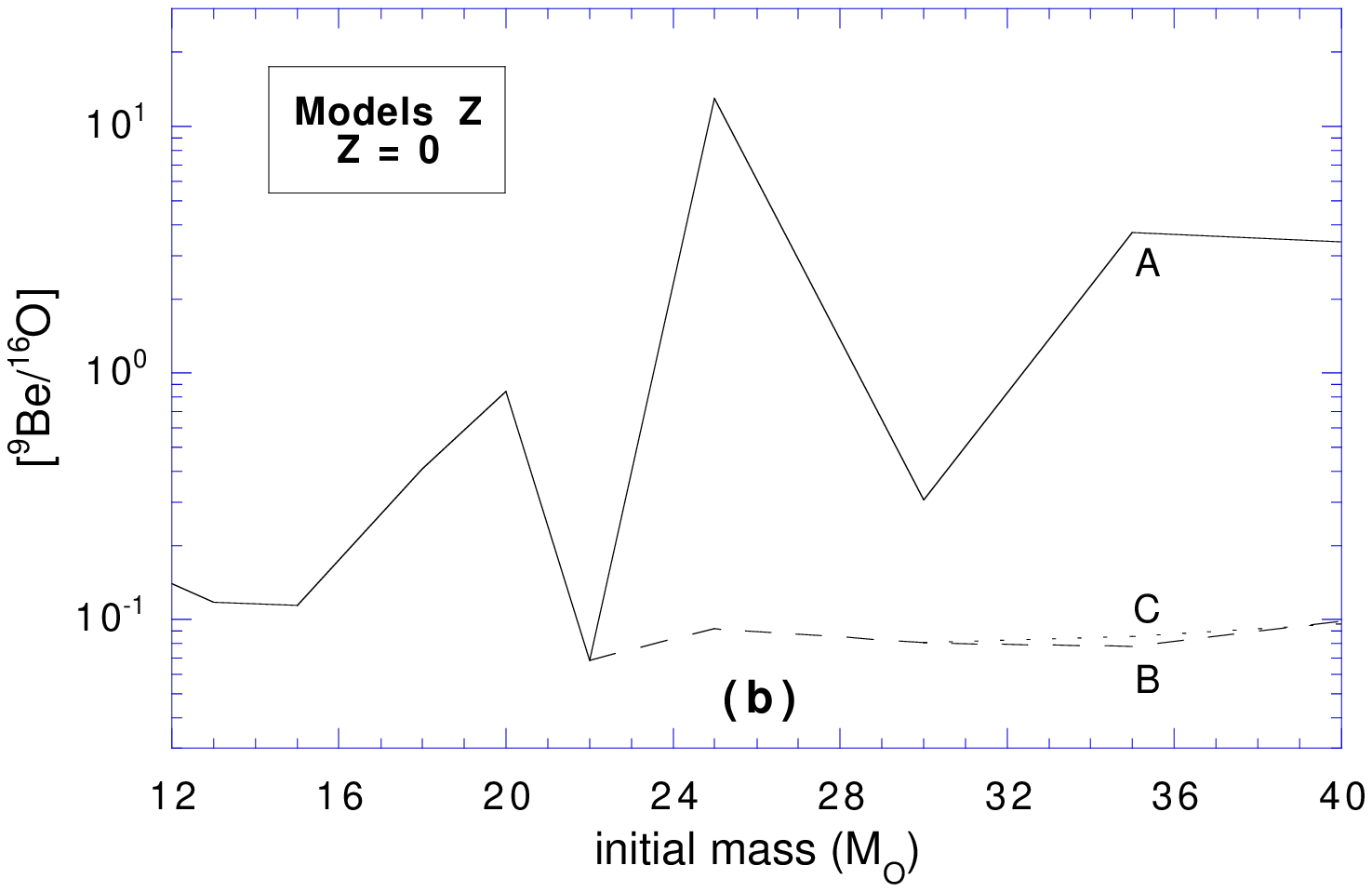, width=8.8cm}}
\centerline{\psfig{file=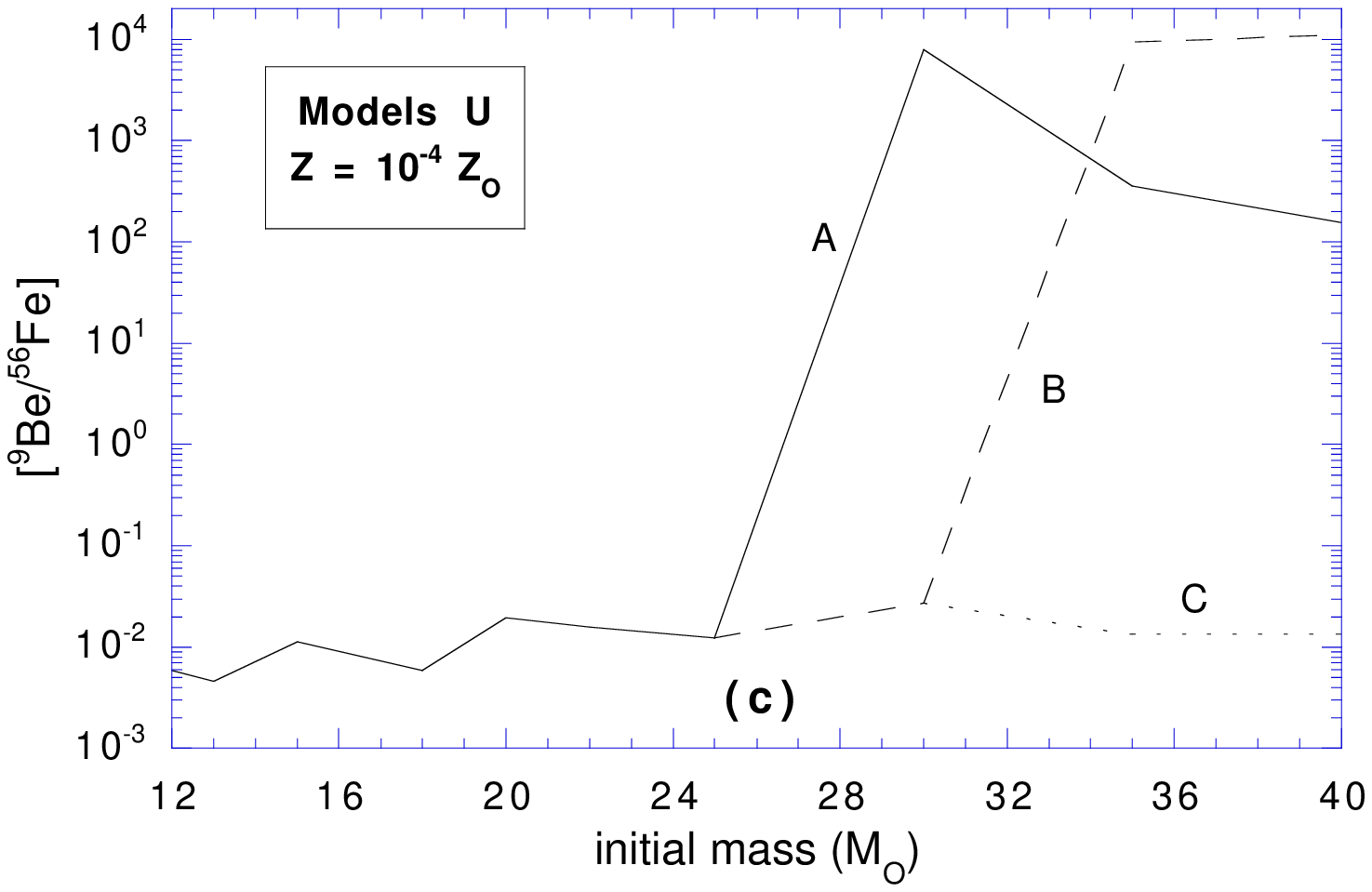, width=8.8cm}
            \hfill
            \psfig{file=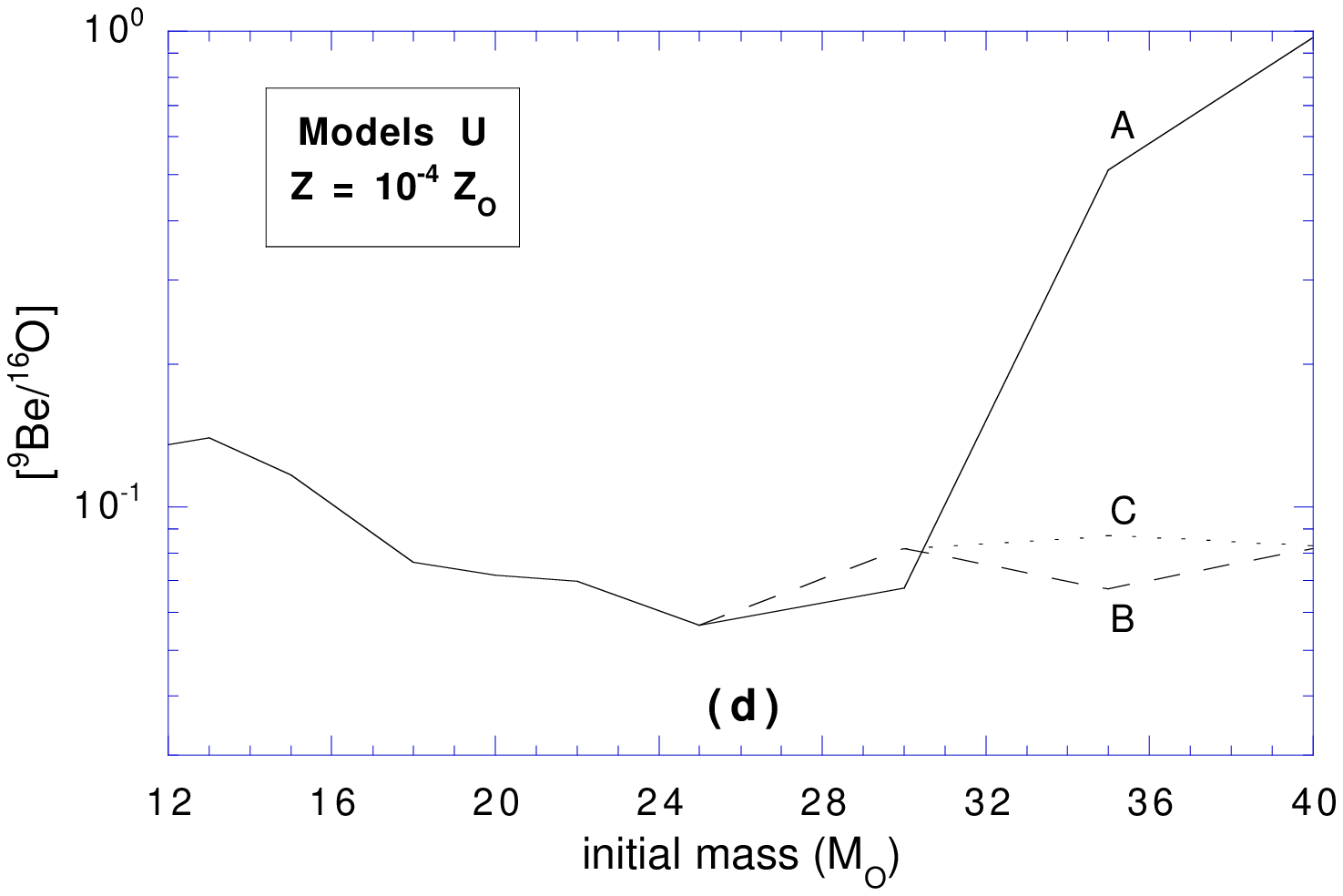, width=8.8cm}}
\centerline{\psfig{file=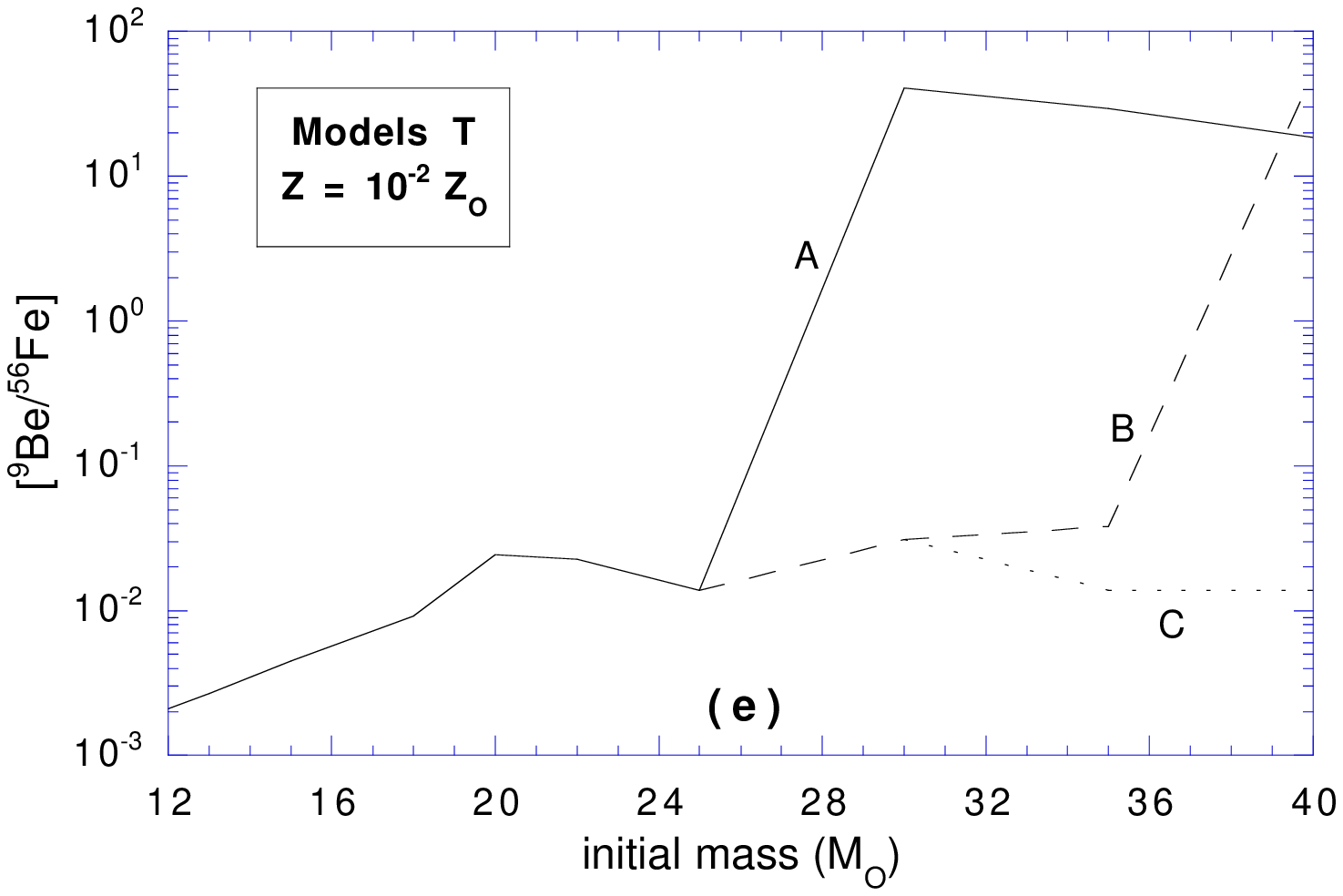, width=8.8cm}
			\hfill
			\psfig{file=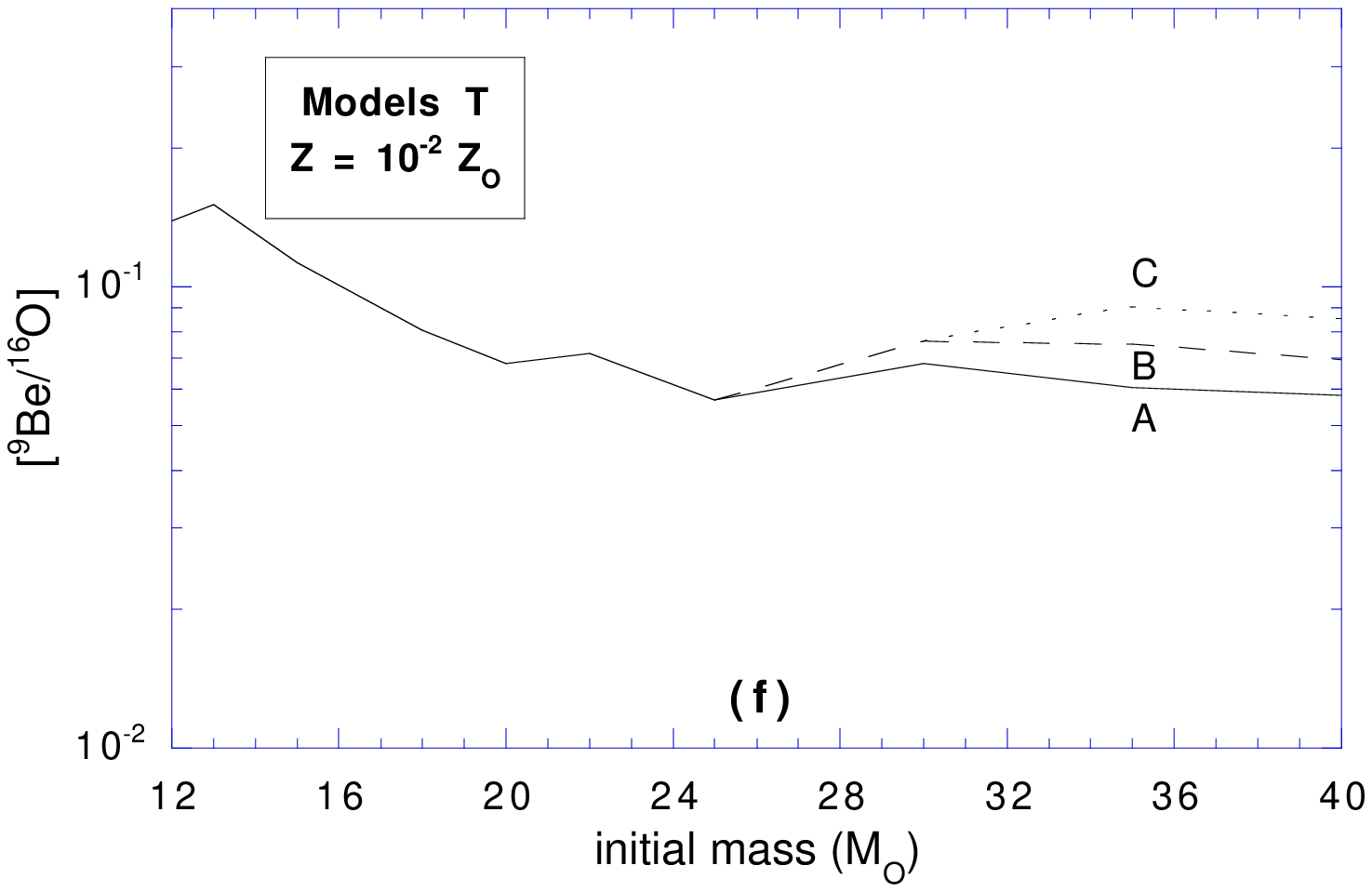, width=8.8cm}} 
\caption{Normalized process~2 [Be/Fe] and [Be/O] yield ratios, as a function of 
the mass of the progenitor.  Models Z, U and T correspond to the 
indicated initial metallicity of the stars.  Models A, B and C 
correspond to different explosion energies (see text and Woosley and 
Weaver, 1995).  The yield ratios are normalized to the value required 
by the observations as explained in the text.}
\label{SNR2_f7}
\end{figure*}

As noted above, however instructive the examination of the spallation 
rates evolution may be, they cannot be directly compared to any 
observational data.  We therefore calculated the (more relevant) 
integrated yields for different models corresponding to initial 
metallicities $Z = 0$ (models Z), $Z = 10^{-4}\,Z_{\odot}$ (models U), 
and $Z = 10^{-2}\,Z_{\odot}$ (models T), and normalized them to the 
expected value, i.e.  to the value required to explain the abundances 
observed in the metal-poor stars.  Consequently, normalized yields 
respectively lower and higher than 1 are equivalent to under- and 
over-production of Be.  A few words of explaination are however 
required as how the normalization is actually performed.  The only 
assumption here is that the Galactic Be evolution is primary relative 
to both Fe and O. This means that the Be/Fe and Be/O ratios are 
approximately constant in metal-poor stars (as is consequently the 
O/Fe ratio).  Then each supernova must lead, on average (over the 
IMF), to the same Be/Fe and Be/O ratios as those observed.  These are 
thus the values we use to normalize our results.  Now, as the Fe and O 
yields calculated by Woosley and Weaver (1995) are different for each 
of their SN models, we applied our normalization model by model and 
obtained the results shown in Fig.\ref{SNR2_f7}, as a function of the 
mass of the SN progenitor, for different initial metallicities.

As discussed earlier, the approximate constancy of the Be/Fe ratio is 
well established observationally, over two orders of magnitude in 
metallicity, from $\mathrm{Fe/H} \la 10^{-3}$ to $10^{-1}$ times the 
solar value.  On the other hand, we still lack similar measurements of 
the Be/O ratio in stars with $\mathrm{O/H} \la 10^{-2}$ times the 
solar value, while the trend at higher metallicity seems to favour a 
slightly increasing Be/O, if one is to believe the recent observations 
by Israelian et~al.  (1998) and Boesgaard et~al.  (1998) (see also 
Fields and Olive, 1999).  To this respect, it might seem that our 
normalization based on the primary behavior of Be is better justified 
for comparison to Fe than to O. In fact, it is just the opposite.  
Indeed, the models we are investigating (processes 1 and~2) predict a 
linear increase of Be as compared to O, whatever the Fe evolution may 
be.  As already noted in Paper~I, Be and Fe actually have no direct 
physical link, as the spallation reactions involve only C and O (and 
in fact mainly Oxygen, as we have shown; see Figs.~\ref{SNR2_f1} 
and~\ref{SNR2_f6}).  Both processes~1 and~2 could therefore account, 
in principle, for any value of the Be/Fe ratio, provided we can choose 
the Iron yield of the SNe (this is however not the case, and even if 
the SN explosion models entail possibly large uncertainties, the claim 
for and use of a constant Be/Fe ratio is in fact justified by the 
observations themselves).  On the contrary, the Be/O ratio is entirely 
determined, at a fundamental level, by the processes we investigate 
here.  A higher mass of Oxygen ejected by the supernova would indeed 
imply a larger Be yield as well, and conversely.

Except for a few `irregular models' which we shall discuss shortly, 
Fig.~\ref{SNR2_f7} shows that the Be yields obtained by process~2 are 
significantly smaller than the required values, by about two orders of 
magnitude when comparison is made with Fe, and roughly one order of 
magnitude when comparison is made with O. This is again in good 
quantitative agreement with the results of Paper~I, so that we confirm 
that the processes considered here cannot be responsible for the 
majority of the Be production in our Galaxy.  This conclusion has 
important implications which have been analysed in Paper~I and will be 
summarized below.  Let us now comment the figures in greater detail.

For each series of explosion models (Z, U and T), corresponding to 
different initial metallicities, Woosley and Weaver (1995) have 
calculated the yields of a number of elements for progenitors of 
different masses ranging from 12 to 40~\Msun.  For the more massive 
progenitors, they found that the yields of Fe, notably, greatly 
depended on the mass-cut, which in their models is directly linked to 
the explosion energy.  For example, a 30~\Msun~model with a `standard' 
explosion energy of $1.2\,10^{51}$~erg ejects virtually no Iron at 
all.  Explosion energies greater than the standard value have 
therefore been explored, leading to higher Fe yields for the most 
massive stars.  We use the same notations as in Woosley and Weaver 
(1995), i.e.  models A, B and C correspond to increasing explosion 
energies of order 1.2, 2 and $2.5\,10^{51}$~ergs, respectively.  In 
fact, the explosion energy has been adjusted for higher mass 
progenitors in an ad hoc way in order to obtain approximately the 
`standard' Fe yield of $\sim 0.1\,\Msun$.  Therefore, passing from 
model A to model B, and finally to model C as the progenitor's mass 
increases, amounts to ensure that the SN yields of both O and Fe do 
not vary in dramatic proportions.  This is the reason why the curves 
for models A, B and C connect so smoothly on Figs.~\ref{SNR2_f7}a-f.  
In particular, it is worth emphasizing that the results which we 
obtain for this `mixed model' (A, then B, then C), are remarkably 
similar whatever the initial metallicity and mass of the progenitor 
may be.  We find in this way $[\mathrm{Be}/\mathrm{Fe}] \sim 0.01$ and 
$[\mathrm{Be}/\mathrm{O}] \sim 0.1$, where the brackets mean that the 
yield ratios have been normalized to the required value as described 
above.

It should be clear, however, that there is no special reason why we 
should increase the explosion energy for the most massive SN 
progenitors.  In fact, the great sensitivity of the Fe yield to the 
explosion energy for these stars mostly means, to our opinion, that 
the SN explosion models are still unable to predict reliable yields 
(especially at the lowest metallicities; see the huge differences 
between the models in Fig.~\ref{SNR2_f7}a).  For instance, if we adopt 
the standard explosion energy (models A), then it is clear from 
Figs.~\ref{SNR2_f7}a,c,e that the observed Be/Fe ratio is very easy to 
reproduce if one assumes that only the most massive stars formed in 
the early Galaxy.  The reason for this success, however, is not that 
the massive stars (indirectly) produce a lot of Be, but rather that 
they produce extremely little Fe.  In this case, then, a serious Fe 
underproduction problem will be encountered by the chemical evolution 
models, so that the high value of the [Be/Fe] should be regarded as 
somewhat artificial, and rather irrelevant to the question of Be 
production in the Galaxy.  Moreover, such a behaviour is not expected 
to be found in the curves showing the Be production as compared to the 
Oxygen.  Indeed, as already alluded to, if a particular SN model 
happens to not eject any substantial amount of O, then it will not 
lead to any significant Be production either, leaving the [Be/O] ratio 
virtually unchanged.  This can be checked on Figs.~\ref{SNR2_f7}b,d,f, 
where all the models are shown to give approximately the same results.  
The only exceptions arise at low metallicity for models A and can be 
easily understood.  In these cases, indeed, the O yield becomes much 
lower than the C yield, so that the Be production is actually 
dominated by spallation reactions involving C. Consequently the Be 
yield is still quite substantial, while the O yield is very low, which 
brings about a situation very similar to that encountered with Fe.

\begin{figure}
\resizebox{\hsize}{!}{\includegraphics{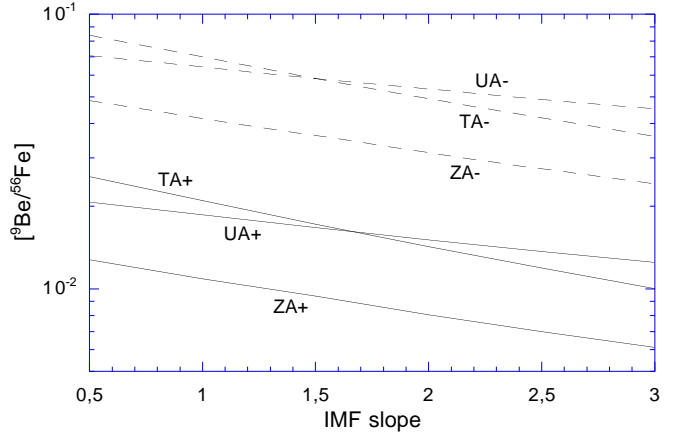}} 
\caption{Normalized Be/Fe ratio calculated from SN models ZA, UA, and 
TA, averaged on the IMF, as a function of the IMF logarithmic slope. 
(Salpeter slope is 2.35). Models labeled with a `+' include adiabatic 
losses; those labeled with a `-' (dashed lines) do not.}
\label{SNR2_f8}
\end{figure}

\begin{figure}
\resizebox{\hsize}{!}{\includegraphics{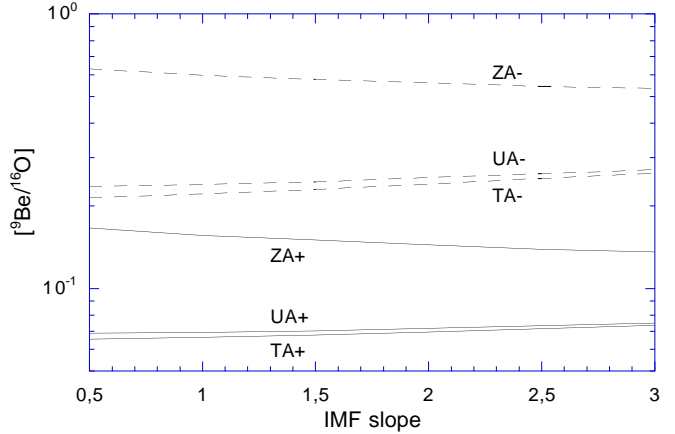}} 
\caption{Normalized Be/O ratio calculated from SN models ZA, UA, and 
TA, averaged on the IMF, as a function of the IMF logarithmic slope. 
(Salpeter slope is 2.35). Models labeled with a `+' include adiabatic 
losses; those labeled with a `-' (dashed lines) do not.}
\label{SNR2_f9}
\end{figure}

However that may be, even if we trust the low (or even extremely low) 
Fe and O yields obtained from models A for high mass progenitors, the 
contribution of these high mass SNe still has to be weighted by their 
frequency among the type~II SNe.  In Figs.~\ref{SNR2_f8} 
and~\ref{SNR2_f9} we show the normalized [Be/Fe] and [Be/O] ratios, 
after averaging over a power-law IMF with logarithmic slope $x$ 
ranging from 0.5 to 3.  This allows us to explore the influence of 
varying the weight of the more efficient high mass stars relatively to 
the lower mass SN progenitors.  A low IMF slope (towards 0.5) strongly 
favours high mass star formation, and is therefore expected to lead to 
a higher [Be/Fe] ratio than a high IMF slope (towards 3).  This 
qualitative behavior is indeed observed on Figs.~\ref{SNR2_f8} 
and~\ref{SNR2_f9}, but it can be seen that the effect is actually 
quite weak, even for such a large range of IMFs.  Note that we used 
`IMFs by number' (of stars), and not `IMFs by mass', so that the 
Salpeter IMF corresponds to $x = 2.35$ in our notations.  This means 
that a slope as low as $x = 0.5$ corresponds to an IMF in which more 
mass is locked in high mass than in low mass stars.  Even for such an 
IMF, the Be/Fe ratio obtained is still less than a few percent of the 
observed value.  Comparing Be to O, it is shown in Fig.~\ref{SNR2_f9} 
that the IMF slope has almost no influence on the normalized [Be/O] 
ratio, which is a consequence of the strong physical link between the 
ejected Oxygen and the Be production, as discussed above.

We have also shown, in Figs.~\ref{SNR2_f8} and~\ref{SNR2_f9}, the 
results obtained without including adiabatic losses (dashed lines).  
Both [Be/Fe] and [Be/O] ratios are then found to be higher by a factor 
of about 3 to 4, which is in good quantitative agreement with the 
analytical calculations of Paper~I (see Fig.~5 there).  This result 
has two simple, but important implications.  First, it points out the 
necessity of including the adiabatic losses in the calculations 
(unless explicitely shown that they do not apply), and therefore of 
using time-dependent models.  Second, it indicates that a model in 
which the EPs do not suffer adiabatic losses has more chance to 
succeed in accounting for the observed amount of Be in the halo stars.

\section{Conclusion}

In conclusion, we have calculated the Be production associated with 
the explosion of a supernova in the ISM, using a time-dependent model, 
and confirmed the results of Parizot and Drury (1999) stating that 
isolated SNe cannot be responsible for the Be observed in the 
metal-poor stars of the Galactic halo.  All the qualitative and 
quantitative features of the two processes investigated (i.e.  
acceleration of particles at the forward and the reverse shocks of an 
isolated supernova) have been found to conform to the analytical 
expectations.  This includes the dependence of the Be yields on the 
ambient density, the evolution of the spallation rates during and 
after the Sedov-like phase of the SNR expansion, and the influence of 
the adiabatic energy losses.

The implications of these results for the Galactic chemical evolution 
of the light elements have been discussed in detail in Paper~I. We 
shall only stress here that it proves very hard for theoretical models 
to produce the required amount of Be (and similarly $^{6}$Li and B) by 
isolated SNe, according to conventional shock acceleration theory.  
Indeed, the processes that we investigated tend to optimize the 
spallation efficiency, in that they either accelerate the freshly 
synthesized C and O or confine the EPs in an environement much richer 
in C and O than the surrounding ISM at this stage of chemical 
evolution.  Shock acceleration efficiencies of order 10 percent are 
also about the maximum that can be expected of \emph{any} acceleration 
process.  Thinking of a process involving more energy than that 
released by a SN and/or a higher concentration of C and O than within 
a SNR is rather challenging.

One promising alternative, however, seems to be a model in which the 
SNe act collectively, rather than individually, as in the processes 
investigated in this paper.  The idea is that most of the massive 
stars in the Galaxy are formed in associations (Melnik and Efremov, 
1995) and generate superbubbles which expand owing to the cumulated 
energy released by several consecutive supernovae.  This energy leads 
to strong magnetic turbulence within the superbubble, which is thought 
to accelerate particles in a very efficient way, according to a 
specific model developed by Bykov and Fleishman (1992).  The 
interesting feature is that the interior of the superbubble is 
enriched by significant amounts of C and O previously ejected by 
stellar winds and SN explosions, so that the accelerated particle 
should have a primary composition (Parizot et al., 1998; Higdon et 
al., 1998; Parizot and Knoedlseder, 1998) and therefore be very 
efficient in producing Be.  Moreover, the average energy imparted to 
the EPs by each supernova is directly related to the explosion energy, 
instead of only the energy in the reverse shock, as in the process~2 
investigated here.  Indeed, either that the particles are accelerated 
directly by the forward shock or that the explosion energy first turns 
into turbulence and a distrubution of weak secondary shocks (this will 
be investigated in a forthcoming paper), the total energy imparted to 
the EPs is expected to be about ten times larger than that assumed for 
process~2 above (say 10~\% of the explosion energy, instead of the 
$\sim 1~\%$ implied by the use of the reverse shock energy).  Further 
considering that the adiabatic losses would not apply in such a case, 
we predict an overall factor of about 10 to 30 on the Be yields, 
depending on the mixing of the ejecta with non enriched ISM within the 
superbubble.  According to the results presented in this paper, this 
would be enough to account for the [Be/O] ratio observed in the 
metal-poor halo stars.

Apart from the problem of light element production in the early 
Galaxy, our calculations have shown that the situation is somewhat 
different whether we compare Be to Fe or O. This obviously indicates 
that the Galactic evolution of Fe and O are mutually inconsistent, if 
one uses the yields of Woosley and Weaver (1995), so that a revision 
of the SN models should be considered.  A similar conclusion has been 
pointed out by Fields and Olive (1999), who observed that these 
theoretical yields cannot reproduce the O/Fe slope measured in the 
abundance diagram.  Since the Be problem is found to be less serious 
when comparison is made with O rather than Fe, we suggest that the Fe 
rather than the O yields may be responsible for the Fe-O problem.  
Further observational and theoretical work are however needed to reach 
a convincing conclusion.

\begin{acknowledgements}
This work was supported by the TMR programme of the European Union 
under contract FMRX-CT98-0168.
\end{acknowledgements}

\end{document}